\documentclass[12pt]{article}
\usepackage{amsmath,amsthm,amssymb,amsfonts,fullpage,verbatim,bm,graphicx,enumerate,epstopdf,lscape,subfigure,multirow}
\usepackage{algorithm,algorithmicx}
\usepackage{algpseudocode}
\usepackage{esvect}
\usepackage{boxedminipage,longtable}
\usepackage[dvipsnames,usenames]{color}
\usepackage[pdftex,letterpaper=true,colorlinks=true,pdfpagemode=none,urlcolor=blue,linkcolor=blue,
   citecolor=blue,pdfstartview=FitH,plainpages=true]{hyperref}
\usepackage{tikz,verbatim,bibentry}
\usepackage{bm,mathrsfs,mathabx}
\usepackage[comma,authoryear]{natbib}
\usepackage{color, colortbl}
\usepackage{dcolumn}

\definecolor{LightCyan}{rgb}{0.88,1,1}
\newtheorem{lemma}{Lemma}
\newtheorem{theorem}{Theorem}
\newtheorem{proposition}{Proposition}

\theoremstyle{definition}
\newtheorem{rmk}{Remark}

\newtheorem{assumption}{Assumption}

\makeatletter
\newcolumntype{X}{>{\rowstyle{\relax}}l}
\newcolumntype{D}[3]{>{\currentrowstyle\DC@{#1}{#2}{#3}}c<{\DC@end}}
\makeatother
\newcommand{\rowstyle}[1]{\gdef\currentrowstyle{#1}}

\newcommand{\utwi}[1]{{\boldsymbol{#1} }}

\newcommand{\bu}{{\utwi{u}}}
\newcommand{\bv}{{\utwi{v}}}
\newcommand{\bw}{{\utwi{w}}}
\newcommand{\bx}{{\utwi{x}}}
\newcommand{\by}{{\utwi{y}}}

\newcommand{\bA}{{\utwi{A}}}
\newcommand{\bB}{{\utwi{B}}}

\newcommand{\bD}{{\utwi{D}}}
\newcommand{\bE}{{\utwi{E}}}
\newcommand{\bF}{{\utwi{F}}}
\newcommand{\bG}{{\utwi{G}}}
\newcommand{\bH}{{\utwi{H}}}
\newcommand{\bI}{{\utwi{I}}}

\newcommand{\bL}{{\utwi{L}}}
\newcommand{\bM}{{\utwi{M}}}

\newcommand{\bO}{{\utwi{O}}}

\newcommand{\bQ}{{\utwi{Q}}}
\newcommand{\bR}{{\utwi{R}}}

\newcommand{\bU}{{\utwi{U}}}
\newcommand{\bV}{{\utwi{V}}}
\newcommand{\bW}{{\utwi{W}}}
\newcommand{\bX}{{\utwi{X}}}
\newcommand{\bY}{{\utwi{Y}}}
\newcommand{\bZ}{{\utwi{Z}}}

\newcommand{\bGamma}{{\utwi{\Gamma}}}

\newcommand{\bSigma}{{\utwi{\Sigma}}}

\newcommand{\balpha}{{\utwi{\alpha}}}

\newcommand{\bgamma}{{\utwi{\gamma}}}

\newcommand{\bzeta}{{\utwi{\zeta}}}

\newcommand{\bpsi}{{\utwi{\psi}}}

\newcommand{\bPhi}{{\utwi{\Phi}}}
\newcommand{\bxi}{{\utwi{\xi}}}

\newcommand{\cF}{{\cal F}}

\newcommand{\cS}{{\cal S}}

\newcommand{\rc}[1]{{\color{blue}{#1}}}

\newcommand{\bel}{\begin{eqnarray}\label}
\newcommand{\eel}{\end{eqnarray}}

\def\mat{\hbox{\rm mat}}
\def\vec{\hbox{\rm vec}}

\def\I{\mathrm{I}}
\def\II{\mathrm{II}}

\def\Cov{{\rm Cov}}
\def\Var{{\rm Var}}
\def\E{\mathbb{E}}
\def\P{\mathbb{P}}
\def\RR{\mathbb{R}}
\def\R{\mathbb{R}}

\def\tr{\mathrm{tr}}

\DeclareMathOperator*{\argmin}{\arg\min}

\begin{document}

\title{\bf Dynamic Matrix Factor Models for High Dimensional Time Series\thanks{\footnotesize{Ruofan Yu is Applied Scientist, Amazon. Email:ruofan@amazon.com. Rong Chen is Professor, Department of
    Statistics, Rutgers University, Piscataway, NJ 08854. E-mail:
    rongchen@stat.rutgers.edu.  Han Xiao is Professor, Department of Statistics, Rutgers University, Piscataway, NJ 08854. E-mail:
    hxiao@stat.rutgers.edu. Yuefeng Han is Assistant Professor, Department of Applied and Computational Mathematics and Statistics, University of Notre Dame, Notre Dame, IN 46556. Email: yuefeng.han@nd.edu.
    Yuefeng Han is the corresponding author.
    Yu's research was done when he was at Rutgers University and was not related to Amazon.  Han's research was supported in part by National Science Foundation grant IIS-1741390. Chen's research was supported in part by National Science Foundation grants DMS-1737857, IIS-1741390, CCF-1934924, DMS-2027855, DMS-2052949 and DMS-2319260. Xiao's research is supported by National Science Foundation grants DMS-2027855, DMS-2052949 and DMS-2319260.}} }

\author{Ruofan Yu$^1$, Rong Chen$^2$, Han Xiao$^2$ and Yuefeng Han$^3$}

\date{$^1$Amazon, USA, $^2$Department of Statistics, Rutgers University, Piscataway, NJ 08854, USA,\\
$^3$Department of Applied and Computational Mathematics and Statistics, University of Notre Dame, Notre
Dame, IN 46556, USA}

\maketitle

\date{}
\maketitle

\begin{abstract}
    Matrix time series, which consist of matrix-valued data observed over time, are prevalent in various fields such as economics, finance, and engineering. Such matrix time series data are often observed in high dimensions. Matrix factor models are employed to reduce the dimensionality of such data, but they lack the capability to make predictions without specified dynamics in the latent factor process. To address this issue, we propose a two-component dynamic matrix factor model that extends the standard matrix factor model by incorporating a matrix autoregressive structure for the low-dimensional latent factor process. This two-component model injects prediction capability to the matrix factor model and provides deeper insights into the dynamics of high-dimensional matrix time series. We present the estimation procedures of the model and their theoretical properties, as well as empirical analysis of the estimation procedures via simulations, and a case study of New York city taxi data, demonstrating the performance and usefulness of the model.
\vspace{0.1in}

\end{abstract}

\section{Introduction}

Tensor data are widely collected in various fields and have become increasingly popular. When the data is collected over time, {one of the modes often represents time, leading to a unique treatment of this temporal dimension.} This type of data is referred to as matrix time series data when the data collected at each time point forms a matrix. Matrix time series data can be found in many applications, such as economics where economic indicators are reported from different countries on a monthly or quarterly basis. These data can be modeled as a matrix time series, where the countries are represented by rows and the economic indicators by columns. In this case, the variables in the same column will be strongly correlated due to reporting from the same country, and the variables in the same row will also be correlated due to the influence of different countries. Thus, it is more reasonable to consider the matrix time series as a whole, preserving the row-wise and column-wise relationships. Another example is a sequence of images, which can be modeled as matrices observed over time. Using matrix models is essential as the spatial structure would be lost if modeled in vector or univariate form.
Recent studies on matrix-valued time series include work by \cite{wang2019} and \cite{chen2022factor}, which analyze factor models for matrix and tensor time series, and studies by \cite{Hoff2015}, \cite{chen2021autoregressive}, and \cite{han2023rr}, which explore autoregressive models in bilinear matrix form.

One of the major challenges in analyzing matrix time series is its high dimensionality. The curse of dimensionality is a notorious problem for high-dimensional data analysis, as it makes it difficult to analyze the data and extract meaningful insights. To overcome this challenge, a common approach is to adopt a factor approach. The factor model describes the co-movement of data with lower-dimensional latent factors. It is based on the idea that high-dimensional data can be reduced to lower-dimensional representations by extracting a small number of underlying factors that capture the most important information in the data.

Factor models are widely used in a variety of fields, including finance, economics, and social sciences. 
{Numerous studies have focused on factor models for time series, with seminal works like \cite{bai2002, bai2003, stock2005} exploring vector time series factor models.}
{Extensions to matrix and tensor factor models have been introduced in \cite{wang2019, chen2022factor, han2020, han2023cp, chen2023statistical, chen2024semi}, among others. In certain studies, the latent factor process is presumed to follow specific dynamic structures, typically autoregressive} \citep{Molenaar1985ADF,tiao1989,engle1995,forni2000,Jasiak2001,hallin2016,fan2013}.
{A critical aspect of building a factor model is to determine the number of factors. This topic has been thoroughly explored in vector factor models, such as \cite{bai2002,lam2012,hallin2007,amengual2007}.}
For matrix and tensor factor models, \cite{han2022rank} proposed information criteria and eigenvalue ratio based methods that use specified penalty functions, {while} \cite{lam2021rank,chen2024rank} proposed a correlation thresholding method without using penalty functions and is free of parameter tuning.

While there have been numerous studies on dynamic factor models for panel data or vector time series, there is still a lack of literature on matrix-valued dynamic factor models. \cite{wang2019} introduces a factor model for matrix-valued time series, but it lacks an assumed dynamic structure for the factor process, making it difficult to predict the factor model without additional specifications. Our goal is to extend the matrix factor model in \cite{wang2019,chen2022factor} by including a dynamic structure for the latent matrix factor process. This approach is similar to dynamic factor models for vector time series in \cite{stock2016}. Specifically, we assume that the matrix factor process follows a matrix autoregressive (MAR) structure as in \cite{Hoff2015, chen2021autoregressive}, represented by $\bF_t = \bA_1\bF_{t-1}\bA_2^{\top} + \bE_t$. The use of this dynamic structure brings two main benefits: (i) It allows for straightforward prediction of the factor model through a two-step computational efficient procedure. (ii) The autoregressive dynamic structure provides a clear and insightful understanding of the hidden dynamics of the factor process. 
It should be noted that our work differs from that of \cite{yuan2023two}, which is based on maximum likelihood estimation and explores a different matrix factor model.

The rest of this paper is organized as follows.
In section 2 we specify the proposed dynamic matrix factor model, with detailed discussion of the motivations using such a model.
In Section 3, we propose a two step estimation procedure for dynamic matrix factor model.
The prediction procedure 
is also discussed.
The theoretical properties of the estimators are presented in Section 4.
Empirical studies are presented in Section 5
to demonstrate the finite sample properties of the estimators. In Section 6, an analysis of a transport network example is shown to demonstrate the usefulness of the model. Section 7 concludes.


\section{Dynamic Matrix Factor Models}\label{section:model}

The proposed Dynamic Matrix Factor Model (DMFM) is a combination of two components, the matrix factor model of \cite{wang2019} and the matrix autoregressive model of \cite{chen2021autoregressive}.
The core innovation lies in its two-component structure, which uniquely combines the strengths of the two models for dimensionality reduction and time series modeling.

\subsection{Model setting}

The Dynamic Matrix Factor Model (DMFM) takes the following two-component form:
\begin{align}
&\bX_t=\lambda \bU_1 \bF_t \bU_2^{\top} + \bE_t, \label{eq:dmfm1}\\
& \bF_t=\bA_1 \bF_{t-1} \bA_2^{\top} +\bxi_t,  \label{eq:dmfm2}
\end{align}
where in the first part \eqref{eq:dmfm1}, $\bX_t \in \R^{d_1\times d_2}$ is the observed matrix time series,
$\bF_t\in \R^{r_1\times r_2}$ is the unobserved latent factor process, with $r_1 \ll d_1$ and $r_2 \ll d_2$. $\bU_1\in \R^{r_1\times d_1}$ and $\bU_2\in \R^{r_2\times d_2}$ are orthonormal loading matrices, and $\lambda$ singles out the signal strength. {In the typical strong factor model setting, $\lambda\asymp \sqrt{d_1 d_2}$.}
We assume the $\bE_t\in \R^{d_1\times d_2}$ is a white noise series so that
$\Cov(\vec(\bE_{t_1}),\vec(\bE_{t_2}))=0$ for $t_1\neq t_2$. Contemporary correlations among the elements of $\bE_t$ are allowed, with $\Cov(\vec(\bE_t)) = (d_1d_2)^{-1} \boldsymbol{\Sigma}_E$.

The second part (\ref{eq:dmfm2}) of DMFM aims to describe the dynamics of factors after dimension reduction by the matrix factor model (\ref{eq:dmfm1}). We assume that the latent factor process $\bF_t$ follows an autoregressive model, with the coefficient matrices $\bA_1\in \R^{r_1\times r_1}$ and $\bA_2\in \R^{r_2\times r_2}$ being the left and right autoregressive coefficients. $\bxi_t\in \R^{r_1\times r_2}$ are innovation series which are assumed to be white noise but also with possible contemporary correlations among its elements, with $\Cov(\vec(\bxi_t)) = (r_1r_2)^{-1}\bSigma_\bxi$.
In addition, we assume that $\bE_t$ and $\bxi_{t+h}$ are uncorrelated for all $h\in \mathbb{Z}$, making $\bE_t$ and $\bF_{t+h}$
uncorrelated for all $h \in \mathbb{Z}$ as well.

The first component \eqref{eq:dmfm1} is a Matrix Factor Model \citep{wang2019}. It projects the temporal dynamics of the high-dimensional matrix time series $\bX_t$ into the temporal dynamics of the low-dimensional latent factor $\bF_t$. This transformation is crucial for computational efficiency and for the extraction of meaningful latent features from the raw data. 
The second component \eqref{eq:dmfm2} is a Matrix Autoregressive Model \citep{chen2021autoregressive} that provides a parametric model for the dynamics of the latent factors. It makes the DMFM a generative model with a relatively simple prediction capability which the matrix factor model \eqref{eq:dmfm1} alone does not have. \cite{wang2024high} considered a low rank tensor AR model and uses a nuclear norm penalty to enforce the low rank structure and optimization algorithms for estimation. It is quite different from our approach.

\begin{rmk}
    The Matrix AR model in \eqref{eq:dmfm2} is a one-term MAR model with autoregressive order of 1. A multi-term AR($p$) model is in the form of
\[
\bF_t=\sum_{i=1}^p\sum_{j=1}^k\bA_{1ij}\bF_{t-i}\bA_{2ij}+\bxi_t.
\]
See more details in \cite{chen2021autoregressive} and \cite{li2023multilinear}. For notation simplicity, all developments in this paper is for the simplest case \eqref{eq:dmfm2}.
\end{rmk}

\begin{rmk}
The second component 
(\ref{eq:dmfm2}) can be replaced by other models. For example, if $r_1$ and $r_2$ are small, a vector AR model can be used for $\vec{(\bF_t)}$. An even simpler approach is to treat each element of $\bF_t$ as an independent univariate time series, and build different linear or nonlinear models for each element series.
\end{rmk}

\begin{rmk}
This matrix model can be extended to its tensor version, dynamic tensor factor time series model, using the tensor factor model introduced in \cite{chen2022factor} and the tensor autoregressive model in \cite{li2023multilinear}. The second part of the model can adopt multi-term autoregressive models or higher autoregressive orders as well.
\end{rmk}

\subsection{Relation to Reduced Rank MAR Model}\label{section:relation to RRMAR}
A reduced rank MAR Model was developed in \cite{han2023rr}, which has a close connection with dynamic matrix factor model. 
The reduced rank MAR Model takes the form
\begin{align}
\bX_t= \bA_{1l}\bA_{1c}^{\top} \bX_{t-1} \bA_{2c}\bA_{2l}^{\top} +\bE_t,  \label{eq:model4}
\end{align}
where  $\bA_{il}$ and $\bA_{ic}$ are both $d_i\times r_i$ full rank matrices. It is essentially the MAR model
$\bX_t=\bA_1 \bX_{t-1} \bA_2^{\top} +\bE_t$, with
$\bA_i = \bA_{il}\bA_{ic}^{\top}$ ($i=1,2$)
being a $d_i\times d_i$ matrix of rank $r_i$.


Let $\bF^*_t = \bA_{1c}^{\top} \bX_{t-1}\bA_{2c}$ and notice that
\[
\bF^*_t = \bA_{1c}^{\top} \bX_{t-1}\bA_{2c}
=\bA_{1c}^{\top}[\bA_{1l}\bA_{1c}^{\top} \bX_{t-2} \bA_{2c}\bA_{2l}^{\top}+\bE_{t-1}]\bA_{2c}
=\bA^*_{1}\bF^*_{t-1}\bA_{2}^{*\top}+ \bA_{1c}^{\top}\bE_{t-1} \bA_{2c},
\]
where $\bA_i^*=\bA_{ic}^\top\bA_{il}$.
Hence model \eqref{eq:model4} becomes
\[
\bX_t= \bA_{1l}\bF_t^*\bA_{2l}^{\top} +\bE_t, \mbox{\ \ }
\bF^*_t =\bA^*_1 \bF^*_{t-1} \bA_2^{*\top} + \bA_{1c}^{\top}\bE_{t-1} \bA_{2c}.
\]
This is very similar to DMFM model in \eqref{eq:dmfm1} and \eqref{eq:dmfm2}, with constraints in the coefficient matrices and shared noise process in $\bE_t$ and $\bA_{1c}^{\top}\bE_{t-1} \bA_{2c}$ in the two components.



Similarly, from \eqref{eq:dmfm1} and \eqref{eq:dmfm2}
we have for the DMFM model
\begin{align*}
& \bX_t =\lambda \bU_1 \bF_t \bU_2^{\top} +\bE_t=\lambda \bU_1 \bA_1 \bF_{t-1} \bA_2^{\top} \bU_2^{\top} + \lambda \bU_1 \bxi_t \bU_2^{\top} +\bE_t\\
&\quad = \bU_1 \bA_1 \bU_1^{\top} (\bX_{t-1}-\bE_{t-1}) \bU_2 \bA_2^{\top} \bU_2^{\top} + \lambda \bU_1 \bxi_t \bU_2^{\top}+\bE_t \\
&\quad = \bU_1 \bA_1 \bU_1^{\top} \bX_{t-1} \bU_2 \bA_2^{\top} \bU_2^{\top} + \lambda \bU_1 \bxi_t \bU_2^{\top}- \bU_1 \bA_1 \bU_1^{\top} \bE_{t-1} \bU_2 \bA_2^{\top} \bU_2^{\top} + \bE_t.
\end{align*}
This is in a MAR model form, with reduced rank coefficient matrices $\bU_i \bA_i \bU_i^{\top}$, and error process
$\lambda \bU_1 \bxi_t \bU_2^{\top}- \bU_1 \bA_1 \bU_1^{\top} \bE_{t-1} \bU_2 \bA_2^{\top} \bU_2^{\top}+\bE_t$. The noise process consists of a moving average process involving both $\bE_t$ and $\bE_{t-1}$, and another independent process involving $\bxi_t$. Even without the noise term $\bE_t$, the model is not equivalent to the reduced rank MAR as the noise
$\lambda \bU_1 \bxi_t \bU_2^{\top}$ is now restricted to the space determined by
the coefficient matrices $\bU_i$.



\section{Estimation and Prediction}

\subsection{Model Formulation}
\label{sec:mod_ref}

Before introducing the estimation and prediction procedures, we revisit the model formulation \eqref{eq:dmfm1} and \eqref{eq:dmfm2} and make an equivalent modification to facilitate the discussion. The latent equation \eqref{eq:dmfm2} should be understood as the generating mechanism of a MAR(1) model with coefficient matrices $\bA_i\ (i=1,2)$ and innovation covariance matrix $\bSigma_\bxi$, which determines the scale of the factor process $\bF_t$. The strength of the factor component is then determined by $\lambda$ in \eqref{eq:dmfm1}. However, the estimation of $\bA_i$ and the subsequent predictions are not affected by the scale of the factor process $\hat\bF_t$ since an autoregressive model is entertained in \eqref{eq:dmfm2}. In other words, if we absorb the strength $\lambda$ into the factors $\bF_t$, both the estimation of $\bA_i$ and the prediction of $\bX_t$ remain unchanged. Therefore, we consider the equivalent formulation of the DMFM model:
\begin{align}
  \bX_t &=\bU_1 \bF_t \bU_2^{\top} + \bE_t, \label{eq:dmfm1'}\\
  \bF_t & =\bA_1 \bF_{t-1} \bA_2^{\top} +\bxi_t,  \label{eq:dmfm2'}
\end{align}
where $\lambda$ is absorbed into $\bF_t$. Note that the scale of $\bxi_t$ is also multiplied by $\lambda$, comparing to \eqref{eq:dmfm2}.

Our subsequent discussion on the estimation and prediction will be based on \eqref{eq:dmfm1'} and \eqref{eq:dmfm2'}. As a result, we will suppress the estimation of $\lambda$ and focus on $\bU_i$ and $\bA_i$. More specifically, after obtaining the two orthonomal loading matrices $\hat\bU_1$ and $\hat\bU_2$, we will treat $\hat\bF_t=\hat\bU_1^\top\bX_t\hat\bU_2$ as the estimated factors and proceed with the estimation of $\bA_i$ using $\hat\bF_t$.

The estimation error in $\hat\bF_t$ involves not only the error in $\hat\bU_i$, but more importantly,  the noise $\bE_t$ in \eqref{eq:dmfm1'}. One of the major contributions of this paper is to propose a lag-2 moment based estimator of $\bA_i$ to alleviate the impact of $\bE_t$, and to introduce a corresponding prediction procedure. To facilitate the discussion, we note that if the $\bU_i$ are known, the ``estimated" factors $\tilde\bF_t:=\bU_1^\top\bX_t\bU_2$
is $\tilde\bF_t=\bF_t+\bU_1^\top\bE_t\bU_2$. We therefore utilize the following MAR model with the measurement error to introduce the idea and method of estimating $\bA_i$.
\begin{align}
  \tilde\bF_t &=\bF_t + \bzeta_t, \label{eq:mar_me1}\\
  \bF_t & =\bA_1 \bF_{t-1} \bA_2^{\top} +\bxi_t,  \label{eq:mar_me2}
\end{align}
It is immediately seen that the preceding model is 
in the state-space form, where $\bF_t$ are latent state variable, and $\tilde\bF_t$ are observed. The link between the DMFM model \eqref{eq:dmfm1'} and \eqref{eq:dmfm2'} and the preceding state space model is that $\bzeta_t=\bU_1^\top\bE_t\bU_2$. We use $\bSigma_\zeta$ to denote the covariance matrix of $\vec(\bzeta_t)$.

We propose a two-stage estimation procedure for the dynamic matrix factor model. In the first stage, we employ matrix factor model estimation to obtain $\hat\bU_i$ and $\hat{\boldsymbol{F}}_t=\hat\bU_1^\top\bX_t\hat\bU_2$. Subsequently, we treat the estimated factor series as observed with measurement error, and use it to estimate the parameters in the matrix autoregressive model. Such a two-stage estimation procedure is commonly used in estimating dynamic factor models \citep{stock2016,Jasiak2001,hallin2016,otto2022approximate}.

For estimating the matrix factor model component \eqref{eq:dmfm1}, we use the iTOPUP and iTIPUP methods, as introduced in \cite{chen2022factor} and \cite{han2020}. These methods are extended version of principal component analysis (PCA) and are designed for matrix and tensor time series. The resulting estimators are denoted as $\hat{\lambda}$, $\hat{\bU}_1$ and $\hat{\bU}_2$, respectively. 

For estimating the MAR coefficient matrices $\bA_i$ in \eqref{eq:dmfm2'}, the PROJ, LSE and MLE estimators in \cite{chen2021autoregressive} can be applied directly, treating
$\hat\bF_t$ as the observation. These estimators are obtained by directly fitting a MAR model on $\hat{\bF}_t$ from the first stage, thus ignoring the estimation error in $\hat\bF_t$ (also see \eqref{eq:mar_me1}). However, the estimation error in $\hat\bF_t$ can be substantial when the signal to noise ratio is at a low level. This is similar to the case of modeling time series observed with measurement errors. Hence, we propose a lagged estimator that reduces the influence of the errors from the factor estimation. To enhance the
estimation of the factors given all the estimated parameters in both parts of DMFM, we also use Kalman Filter to obtain more accurate estimate of the factor process under the state-space formulation \eqref{eq:mar_me1} and \eqref{eq:mar_me2}.

The estimation accuracy is driven by the signal-to-noise ratio (SNR) of the model. There are in fact two relevant SNRs, one related to the estimation of the loadings $\bU_i$, and one related to the estimation of $\bA_i$. Since we are adopting the existing method of \cite{han2020} for the estimation of $\bU_i$, we will only introduce the SNR relevant to the estimation of $\bA_i$. In view of \eqref{eq:mar_me1}, we define
\begin{align}\label{eq:snr}
\text{SNR}:=\sqrt{\frac{\E \|  \bF_{t}  \|^2_{\rm F}} {\E\|\bzeta_t\|^2_{\rm F}}} = \sqrt{\frac{\E \|  \bF_{t}  \|^2_{\rm F}} {\E \|  \bU_1^\top\bE_{t}\bU_2  \|^2_{\rm F}}}.
\end{align}

\subsection{Factor estimation and measurement error}
The first step is to determine the rank of the factor matrix. To achieve this, we utilize the method proposed in \cite{han2022rank}, which suggests either the Bayesian Information Criterion (BIC) or the eigenvalue ratio (ER) method for estimating the number of factors. Once the ranks $r_1$ and $r_2$ are determined, we can estimate the loading matrices $\hat{\boldsymbol{U}}_1$ and $\hat{\boldsymbol{U}}_2$, as well as the estimated latent factor process $\hat{\boldsymbol{F}}_t$, using the iTIPUP or iTOPUP matrix factor estimation methods. It is important to note that the estimated $\hat{\boldsymbol{U}}_1$, $\hat{\boldsymbol{U}}_2$, and $\hat{\boldsymbol{F}}_t$ are not unique; however, any choice of these estimates will not impact the prediction results. In the following sections, we will delve deeper into the identification problem.
The estimation error in $\hat{\bF}_t$ may have significant impact on the second stage estimation. Without considering the dynamics in the factor process, \cite{han2020} established that the error rate of $\hat{\bF}_t$ is of the order of $O_{\P} \left(\frac{\sigma\sqrt{d_{\max}}}{\lambda\sqrt{T}}+\frac{\sigma}{\lambda} \right)$; {see also \eqref{eq:prop:factor}}. When the signal to noise ratio is high,
the error is small enough to obtain accurate estimation in the second stage. When the signal to noise ratio is not sufficiently high,
we introduce a (higher) lagged estimator to alleviate the influence of estimation error from the estimation of the factors.

\subsection{Estimation of the coefficient matrices in \eqref{eq:dmfm2}} \label{sec:standard}

Here we propose two moment estimators of $\bA_i$ in \eqref{eq:dmfm2}.
Although the second estimator is designed to deal with
un-ignorable measurement errors in
$\hat{\bF}_t$, in this section we simply assume the first stage
factor estimation produces the factor process without error, and develop the
estimator of the coefficient matrices in \eqref{eq:dmfm2} with the observed true $\bF_t$ process. The consequences of the measurement error will become apparent later in our empirical and theoretic investigation of the estimators.

\subsubsection{Lag-1 Moment Estimator (LSE)}\label{sec:lse}


We consider lag-1 Yule-Walker estimator in the form of
\begin{align}
 \min\limits_{\bA_1,\bA_2}\| (\bA_2 \otimes \bA_1- \hat{\bGamma}_1\hat{\bGamma}_0^{-1}) \hat{\bGamma}_0^{1/2}\|_{\rm F}^2, \label{eq:YW1}
\end{align}
where $\hat\bGamma_0$ and $\hat\bGamma_1$ are the lag-0 and lag-1 sample autocovariance matrices of the vectorized factor series $\vec(\bF_t)$, i.e. $\hat\bGamma_k=T^{-1}\sum_{t=k+1}^T \vec(\bF_t)\vec(\bF_{t-k})^\top$, for $k\geq 0$. Note that $\hat{\bGamma}_1\hat{\bGamma}_0^{-1}$ is the Yule-Walker estimator of the coefficient matrix $\bPhi$ in the VAR(1) model $\vec(\bF_t)=\bPhi\vec(\bF_{t-1})+\vec(\bxi_t)$. To estimate $\bA_i$ in the MAR model \eqref{eq:dmfm2}, a straightforward approach is to project $\hat{\bGamma}_1\hat{\bGamma}_0^{-1}$ to the cone of Kronecker products through
\begin{equation}\label{eq:proj_est}
    \min\limits_{\bA_1,\bA_2}\| \bA_2 \otimes \bA_1- \hat{\bGamma}_1\hat{\bGamma}_0^{-1}\|_{\rm F}^2.
\end{equation}
The estimator in \eqref{eq:YW1} can be viewed as a projection weighted by $\hat\bGamma_0^{1/2}$.
This weighting scheme is due to the proposition below.

\begin{proposition}\label{prop1:lag-1 yule-walker}
Assuming $\bF_t$ and $\bxi_{t+h}$ are uncorrelated for $h\in \mathbb{Z}$, and the innovations series $\bxi_t$ are i.i.d. with mean zero and variance $\sigma^2$, then
\[
 T^{1/2}\cdot \vec(\hat{\bGamma}_1\hat{\bGamma}_0^{-1}-\bA_2 \otimes \bA_1)\Rightarrow N(0, \sigma^2\bGamma_0^{-1}\otimes\bI),
\]
where $\bGamma_0$ is covariance matrix
of the vectorized factor series $\vec(\bF_t)$.
\end{proposition}

\noindent
The proof is given in Appendix \ref{append:proof}. Proposition~\ref{prop1:lag-1 yule-walker} shows that asymptotic covariance matrix of each row of $\hat{\bGamma}_1\hat{\bGamma}_0^{-1}$ is proportional to $\bGamma_0$, justifying the projection weighted by $\hat\bGamma_0^{1/2}$ in \eqref{eq:YW1}.

A direct calculation shows that \eqref{eq:YW1} is equivalent to
the least squares problem
\begin{align} \label{eq:iterated_ls}
\min\limits_{\bA_1,\bA_2} \sum\limits_{t=2}^T \| \bF_t - \bA_1\bF_{t-1}\bA_2^{\top} \|_{\rm F}^2,
\end{align}
which is considered in \cite{chen2021autoregressive}. Specifically,
the solution of \eqref{eq:iterated_ls} can be found by iteratively updating
\begin{align*}
 \bA_2 &\leftarrow \left(\sum_{t=2}^T \bF_t^{\top} \bA_1 \bF_{t-1} \right) \left(\sum_{t=2}^T \bF_t^{\top} \bA_1^{\top}\bA_1 \bF_{t-1} \right)^{-1}, \\
 \bA_1 &\leftarrow \left(\sum_{t=2}^T \bF_t \bA_2 \bF_{t-1}^{\top} \right) \left(\sum_{t=2}^T \bF_{t-1} \bA_2\bA_2^{\top} \bF_{t}^{\top} \right)^{-1},
\end{align*}
until convergence.
To initialize, one can use the projection estimator in \eqref{eq:proj_est}.

For simplicity, we will refer to the solution of \eqref{eq:iterated_ls} (equivalently \eqref{eq:YW1}) as {\bf LSE}. \cite{chen2021autoregressive} considers both the LSE and the projection estimator \eqref{eq:proj_est}, and has shown empirically that the LSE has a superior performance.

\subsubsection{Lag-2 Moment Estimator (L2E)} \label{sec:lagged}

Similar to the lag-1 Yule-Walker estimator in \eqref{eq:YW1}, 
we can estimate $\bA_1$ and $\bA_2$ by solving the optimization problem
\begin{align}\label{eq:optim2}
    \min\limits_{\bA_1,\bA_2} \| (\bA_2 \otimes \bA_1 - \hat{\bGamma}_2 \hat{\bGamma}_1^{-1}) (\hat{\bGamma}_1 \hat{\bGamma}_0^{-1} \hat{\bGamma}_1^{\top})^{1/2} \|_{\rm F}^2.
\end{align}
The $\hat{\bGamma}_2 \hat{\bGamma}_1^{-1}$ can be viewed as the lag-2 Yule-Walker estimator of the VAR model for $\vec(\bF_t)$. Similar to the LSE, the estimator in \eqref{eq:optim2} is obtained by projecting $\hat{\bGamma}_2 \hat{\bGamma}_1^{-1}$ to the cone of Kronecker products using the weighting matrix $(\hat{\bGamma}_1 \hat{\bGamma}_0^{-1} \hat{\bGamma}_1^{\top})^{1/2}$. The choice of the weighting matrix is motivated by the following proposition.

\begin{proposition}\label{prop2:lag-2 yule-walker}
    Assuming $\bF_t$ and $\bxi_{t+h}$ are uncorrelated for $h\in \mathbb{Z}$, and the innovations series $\bxi_t$ are i.i.d. with mean zero and variance $\sigma^2$, then
    \[
 T^{1/2} \cdot \vec(\hat{\bGamma}_2\hat{\bGamma}_1^{-1}-\bA_2 \otimes \bA_1)\Rightarrow N(0, \sigma^2(\bGamma_1 \bGamma_0^{-1}\bGamma_1^\top)^{-1}\otimes\bI),
\]
where $\bGamma_k$ is the lag-$k$ autocovariance matrix of the vectorized series $\vec(\bF_t)$, for $k=0,1,2$.
\end{proposition}

\noindent
The proof is given in Appendix \ref{append:proof}. For simplicity, we will refer to the estimator in \eqref{eq:optim2} as the {\bf Lag-2 Estimator (L2E)}.

The L2E is designed to handle the measurement error in the observed $\bF_t$, which naturally arises for the DMFM, since the factors are estimated as $\hat\bF_t=\hat\bU_1^\top\bX_t\hat\bU_2=\hat\bU_1^\top\bU_1\bF_t\bU_2^\top\hat\bU_2+\hat\bU_1^\top\bE_t\hat\bU_2$, where $\hat\bU_1^\top\bE_t\hat\bU_2$ becomes the measurement error.
To fix the idea, we consider the MAR model with measurement error \eqref{eq:mar_me1} and \eqref{eq:mar_me2}, where $\tilde\bF_t$ are assumed to be observed.
If the LSE \eqref{eq:YW1} is to be applied to the $\tilde\bF_t$ directly, one needs to calculate $\hat\bGamma_1 \hat\bGamma_0^{-1}$ first, where the sample autocovariance matrices $\hat\bGamma_k$ are calculated for the observed $\tilde\bF_t$. Recall that $\bSigma_\bzeta$ is the covariance matrice of $\vec(\bzeta_t)$. It holds that $\hat\bGamma_0 \asymp \bGamma_0 + \bSigma_\bzeta + O_p(T^{-1/2})$, where $\bGamma_0:=\Var(\vec(\bF_t))$. The bias $\bSigma_\bzeta$ will be translated into the LSE of $\bA_i$ and will only be negligible if $\bSigma_\bzeta$ is of an order smaller than $T^{-1/2}$ (see Theorem~\ref{thm:a} and \ref{thm:b} for more detailed and precise theoretical analysis).

We propose the L2E to alleviate the impact of the measurement error. It is also inspired by the estimator in \cite{hannan1963} that leverages the lagged moments beyond lags 0 and 1.
It effectively minimizes the impact of potential measurement errors. More specifically, for model \eqref{eq:mar_me1} and \eqref{eq:mar_me2}, due to the whiteness of $\bzeta_t$, $\hat\bGamma_k \asymp \bGamma_k +  O_p(T^{-1/2})$ for both $k=1,2$, and the L2E will be asymptotically unbiased as long as $\bSigma_\bzeta=o(1)$ (see Theorem~\ref{thm:lagb}).

The optimization problem in \eqref{eq:optim2} can be solved
as follows.
Let $\bY = \hat\bGamma_2 \hat\bGamma_1^{-1} (\hat\bGamma_1 \hat\bGamma_0^{-1} \hat\bGamma_1^\top)^{1/2}$ and $\bV = (\hat\bGamma_1 \hat\bGamma_0^{-1} \hat\bGamma_1^\top)^{1/2}$. And let $\by_i$ and $\bv_i$ be the $i$-th column of matrix $\bY$ and $\bV$, then \eqref{eq:optim2} becomes equivalent to
\begin{align*}
\min\limits_{\bA_1,\bA_2}\sum\limits_{i=1}^{r_1 r_2} \| \by_i - \bA_2\otimes\bA_1 \bv_i \|_{\rm F}^2 =  \min\limits_{\bA_1,\bA_2}
 \sum\limits_{i=1}^{r_1 r_2} \| \mat(\by_i) - \bA_1 \mat(\bv_i) \bA_2^{\top} \|_{\rm F}^2.  \end{align*}
Now the optimization problem in \eqref{eq:optim2}  can be solved by the iterated least squares method introduced in Section~\ref{sec:lse} similarly.

\subsubsection{Improved factor estimation using Kalman Filter}
\label{sec:kf}

The first step of the estimation produces estimated factor process $\hat{\bF}_t$, using only the factor model component
\eqref{eq:dmfm1}. Since the second component \eqref{eq:dmfm2} also provides
useful information about $\bF_t$, it would be desirable to utilize the full model in the estimation. Given the estimated $\bU_1$, $\bU_2$, $\bA_1$ and $\bA_2$, DMFM becomes a linear state space model with the latent factor process $\bF_t$ as the state and $\bX_t$ as the observation. Hence Kalman filter and smoother \citep{kalman1960} can be used to provide an improved estimator of $\bF_t$. However, using the full DMFM as the state-space model for filtering would involve very high dimensional estimated loading matrices and the estimated covariance matrices of $\vec(\bE_t)$. The estimation errors make the system unstable. Here we use a simplified
approximation by assuming the estimated factor process from the first step is the
true factor process with independent measurement error. Such an approximation
allows us to work with a much simpler
state-space model of low dimensions.


Specifically, we assume
the following vector version of the state space model \eqref{eq:mar_me1} and \eqref{eq:mar_me2}:
\begin{align}
&\vec(\tilde{\boldsymbol{F}}_t) =\vec(\bF_t) +\vec(\bzeta_t), \label{eq:state_space2} \\
&\vec(\bF_{t}) = \bPhi\vec(\bF_{t-1}) +\vec(\bxi_t), \label{eq:state_space1}
\end{align}
where $\bF_{t}$ is the underlying true factor series, and $\bPhi = \bA_2 \otimes \bA_1$ is the coefficient matrix of the VAR model \eqref{eq:state_space1}. Recall that the innovations $\bxi_t$ 
and $\bzeta_t$ have covariance structures
${\rm Cov}(\vec(\bxi_t)) = \bSigma_\bxi$
and ${\rm Cov}(\vec(\bzeta_t)) = \boldsymbol{\Sigma}_{\bzeta}$ respectively. Equation (\ref{eq:state_space1}) is the state equation with $\vec(\bF_t)$ as the state,
and (\ref{eq:state_space2}) is the observation equation, with $\vec(\tilde{\bF_t})$ as the observations.

To estimate the covariance matrices $\bSigma_\bxi$ and $\bSigma_\bzeta$ needed for filtering, we obtain the sample ``residual''
$\bW_t =\vec(\tilde{\bF}_t) - \bPhi\vec(\tilde{\bF}_{t-1}) =\vec(\bzeta_t) - \bPhi\vec(\bzeta_{t-1}) +\vec(\bxi_{t})$. Since
\[
\bG_0 := \Cov(\bW_t) = \bSigma_\bxi + \bPhi \boldsymbol{\Sigma}_\bzeta\bPhi^{\top} + \boldsymbol{\Sigma}_\zeta, \mbox{\ and \ }
\bG_1 := \Cov(\bW_t,\bW_{t-1})=
-\bPhi \boldsymbol{\Sigma}_\bzeta,
\]
we have
\[
\boldsymbol{\Sigma}_\bzeta = -\bPhi^{-1} \bG_1,
\mbox{\ and \ }
\bSigma_\bxi = \bG_0 - \boldsymbol{\Sigma}_\bzeta- \bPhi \boldsymbol{\Sigma}_\bzeta \bPhi^{\top}.
\]
In practice, 
we let the estimated factors $\hat\bF_t$ to play the roles of $\tilde\bF_t$, and use the L2E $\hat\bPhi=\hat\bA_2\otimes\hat\bA_1$ to calculate the $\hat\bW_t =\vec(\hat{\bF}_t) - \hat\bPhi\vec(\hat{\bF}_{t-1})$, from which the sample autocovariance matrices $\hat\bG_0$ and $\hat\bG_1$ are obtained. A direct moment estimator of $\bSigma_\bzeta$ would be $-\hat\bPhi^{-1}\hat\bG_1$ but it may not be symmetric and positive semi-definite as a covariance matrix needs to be. Hence
we project $-\left(\hat\bPhi^{-1}\hat\bG_1+\hat\bG_1^\top\hat\bPhi^{-\top}\right)/2$ to the cone of positive semi-definite matrices. Specifically, let $-\left(\hat\bPhi^{-1}\hat\bG_1+\hat\bG_1^\top\hat\bPhi^{-\top}\right)/2=\hat\bQ\hat\bD\hat\bQ'$ be the eigenvalue decomposition, we take
\begin{equation}\label{eq:sig_err}
    \hat\bSigma_\bzeta := \hat\bQ\hat\bD_+\hat\bQ',
\end{equation}
where $\hat\bD_+$ is the diagonal matrix obtained by thresholding the diagonal entries of $\hat\bD$ from below by zero. The direct moment estimator of $\bSigma_\bxi$ is then given by $\hat\bSigma_\bxi = \hat\bG_0 - \hat{\boldsymbol{\Sigma}}_\bzeta- \hat\bPhi \hat{\boldsymbol{\Sigma}}_\bzeta \hat\bPhi^{\top}$, which again is projected to the cone of positive semi-definite matrices to generate the final estimator $\hat\bSigma_\bxi$.

Once $\hat\bPhi$, $\hat\bSigma_\bxi$ and $\hat\bSigma_\bzeta$ are obtained, the implementation of the Kalman filter based on \eqref{eq:state_space1} and \eqref{eq:state_space2} is standard, so we omit the details here. We remark that the projection step may produce singular $\hat\bSigma_\bxi$ and $\hat\bSigma_\bzeta$, although they are both positive semi-definite. In the implementation, we always use the Moore-Penrose inverse if a singular covariance matrix is encountered at any step of the algorithm.

\subsection{Prediction}
\label{sec:pred}
Denote by $\mathcal{F}_t$ the set of observations $\{\bX_1,\cdots, \bX_t\}$ up to time $t$, and by $\bX_t(1)$ and $\bF_t(1)$ the best linear predictor of $\bX_{t+1}$ and $\bF_{t+1}$ based on $\cF_t$. Under the DMFM model \eqref{eq:dmfm1'} and \eqref{eq:dmfm2'},
\begin{equation}\label{eq:pred0}
    \bX_t(1) = \bU_1\bF_t(1)\bU_2^\top =
    \bU_1\bA_1\bF_t(0)\bA_2^\top\bU_2^\top,
\end{equation}
where $\bF_t(0)$ is the best linear prediction of $\bF_t$ based on $\cF_t$.
To generate the prediction based on the estimated model, a straightforward approach is to plug in the estimates $\hat\bU_i$, $\hat\bA_i$ and $\hat\bF_t$ into \eqref{eq:pred0}. However, the behavior of such a plug-in prediction is subtle, as we will elaborate below. First of all, as to be shown in both Section~\ref{sec:theory} and Section~\ref{sec:sim_est}, when the SNR is relatively small, the L2E has better performance than LSE. However, even when $\hat\bA_i^{\rm (L2E)}$ are closer to the true $\bA_i$ than $\hat\bA_i^{\rm (LSE)}$, a direct plug-in of $\hat\bA_i^{\rm (L2E)}$ into \eqref{eq:pred0} will not lead to a better prediction, because the prediction also involves $\hat\bF_t$ as the prediction of the latent $\bF_t(0)$ by $\cF_t$, but $\hat\bF_t$ is obtained using only \eqref{eq:dmfm1'} without using the dynamics in \eqref{eq:dmfm2'}. It can be shown easily that $\bA_1^{(\rm LSE)}$ and $\bA_2^{(\rm LSE)}$ attempt to minimize the square linear prediction error of $\bF_{t+1}$ by $\hat\bF_{t}$, and hence that of $\bX_{t+1}$ by $\hat\bF_t$, if $\bU_1$ and $\bU_2$ are known. Therefore, the plug-in prediction \eqref{eq:pred0} using LSE is preferred to using L2E, even when the SNR is small, though L2E is the preferred estimator for $\bA_1$ and $\bA_2$ in such cases. Therefore, if a plug-in type prediction is entertained, one should always use the LSE:
\begin{equation}
    \label{eq:pred_plug}
    \hat\bX_t(1) = \hat\bU_1\hat\bA_1^{\rm (LSE)}\hat\bF_t\hat\bA_2^{(\rm LSE)\top}\hat\bU_2^\top.
\end{equation}

However, the proposed L2E can be effectively used to enhance the prediction performance using the underlying state-space structure of the process, instead of the simple plug-in prediction.
Specifically, we propose to use the Kalman filter in Section~\ref{sec:kf} to obtain the best linear estimation 
of $\bF_t$ based on $\{\hat\bF_1,\ldots,\hat\bF_t\}$
as an estimate of $\bF_t(0)$, and then obtain $\bX_t(1)$ using \eqref{eq:pred0}. In practice, the Kalman filter is applied to $\{\hat\bF_t\}$ with the estimated $\hat\bA_i^{\rm (L2E)}$, $\hat\bSigma_\bxi$ and $\hat\bSigma_\bzeta$ to get $\hat\bF_t(0)$. The prediction of $\bX_{t+1}$ is then given by
\begin{equation}\label{eq:pred_kf}
    \hat\bX_t(1) = \hat\bU_1\hat\bA_1^{\rm (L2E)}\hat\bF_t(0)\left[\hat\bA_2^{\rm (L2E)}\right]^\top\hat\bU_2^\top.
\end{equation}
It is critical in practice to decide whether the Kalman filter based prediction \eqref{eq:pred_kf} or the plug-in prediction \eqref{eq:pred_plug} should be adopted, which in turn depends on the level of the SNR. There is some space for a more thorough and detailed analysis on the choice between LSE and L2E, using potentially certain information criterion or bootstrap procedure. Due to the scope of the current paper, we will leave it to the future work. Here we propose a simple rule of thumb based on heuristics. Note that the estimate $\hat\bSigma_\bzeta$ in \eqref{eq:sig_err} is the estimated covariance matrix of the measurement error in $\hat\bF_t$. If the measurement error is significantly large, there is a large chance that $\hat\bSigma_\bzeta$ is strictly positive definite. Therefore, we would use the Kalman filter based prediction \eqref{eq:pred_kf} if the smallest eigenvalue of $\hat\bSigma_\bzeta$ is positive and use the plug-in prediction \eqref{eq:pred_plug} otherwise.

\section{Theoretical properties}
\label{sec:theory}

In this section we present some theoretical properties of the proposed two-stage estimation procedures. To single out the impact of the SNR for the estimation of both $\bU_i$ and $\bA_i$, we return in this section the original model formulation \eqref{eq:dmfm1} and \eqref{eq:dmfm2}, where $\lambda$ is made explicit.
We introduce some notations first. Let $d_{\max}=\max\{d_1,d_2\}$, $\otimes$ be the Kronecker product. The matrix Frobenius norm is defined as $\|\bA\|_{\rm F} = (\sum_{ij} a_{ij}^2)^{1/2}$.
Define the spectral norm as
$$ \|\bA\|_{\rm S} =  \max_{\|\bx\|_2=1,\|\by\|_2= 1} \|\bx^{\top} \bA \by\|_2.$$

To establish the consistency of the proposed procedures, we impose
the following assumptions.

\begin{assumption}\label{asmp:error}
The idiosyncratic noise process $\bE_t$ are independent Gaussian matrices. In addition, there exists some constant $\sigma>0$, such that
\begin{equation*}
\E (\bu^{\top} \text{vec}(\bE_t))^2\le \sigma^2 \|\bu\|_2^2, \quad \forall\,\bu\in\RR^d.
\end{equation*}
\end{assumption}

\begin{assumption}\label{asmp:mar-noise}
The innovation matrices of the latent MAR process, $\bxi_t$ are i.i.d. with mean zero and finite second moments for each elements, and absolutely continuous with respect to Lebesque measure.
\end{assumption}

Assumption \ref{asmp:error} is adopted from \cite{chen2022factor} and \cite{han2020} on tensor factor models, which also corresponds to the white noise assumption of \cite{lam2011,lam2012}. Different from the assumptions in \cite{chen2023statistical}, it allows substantial contemporaneous correlation among the entries of $\bE_t$. Note that the normality assumption, which ensures fast convergence rates in our analysis, is imposed for technical convenience. In fact it can be extended to the sub-Gaussian condition, and still maintains the same convergence rates of the factor loading matrices as those presented in \cite{chen2022factor,han2020}. Assumption \ref{asmp:mar-noise} is a standard condition for matrix autoregressive models.

\cite{han2020} proposed iterative procedures, iTOPUP and iTIPUP, to estimate the factor loading matrices of matrix/tensor factor models. When the ranks $r_k$ and time lag $h_0$ are fixed, both methods delivered convergence rate $\| \widehat\bU_k\widehat\bU_k^{\top} -\bU_k \bU_k^{\top} \|_{\rm S} =O_{\P}(\sigma d_{\max}^{1/2}\lambda^{-1}T^{-1/2}),k=1, 2$, which also matches the statistical lower bound. Using their iterative procedures, we can further establish the theoretical properties of the estimated latent factor process $\widehat \bF_t$ in \eqref{eq:dmfm1}.

\begin{proposition}\label{prop:factor}
Suppose Assumptions \ref{asmp:error} and \ref{asmp:mar-noise} hold. Assume $r_1,r_2$ are fixed. Let $\widehat\bF_t=\widehat\bU_1^{\top} \bX_t \widehat\bU_2/\lambda$ be the estimated factors using iTOPUP or iTIPUP in \cite{han2020}.
Then {there exit rotation matrices $\bR_1, \bR_2$ such that}
\begin{align}\label{eq:prop:factor}
\left\| \widehat\bF_t - \bR_1 \bF_t \bR_2^\top \right\|_{\rm S} =O_{\P} \left(\frac{\sigma\sqrt{d_{\max}}}{\lambda\sqrt{T}}+\frac{\sigma}{\lambda} \right),
\end{align}
and
\begin{align}
&\frac{1}{T-h}\sum_{t=h+1}^T  \vec(\widehat \bF_{t}) \vec(\widehat \bF_{t-h})^\top  - \frac{1}{T-h}\sum_{t=h+1}^T  \vec(\bR_1 \bF_{t} \bR_2^\top) \vec(\bR_1\bF_{t-h} \bR_2^\top)^\top  =  O_{\P} \left(\frac{\sigma\sqrt{d_{\max}}}{\lambda\sqrt{T}} \right), \label{eq:prop:factor1} \\
&\frac{1}{T}\sum_{t=1}^T  \vec(\widehat \bF_{t}) \vec(\widehat \bF_{t})^\top  - \frac{1}{T}\sum_{t=1}^T  \vec(\bR_1 \bF_{t} \bR_2^\top) \vec(\bR_1\bF_{t} \bR_2^\top)^\top  =  O_{\P} \left(\frac{\sigma\sqrt{d_{\max}}}{\lambda\sqrt{T}} +\frac{\sigma^2}{\lambda^2} \right), \label{eq:prop:factor2}
\end{align}
for all $1\le h\le T/4$.
\end{proposition}

The proposition specifies the convergence rate for the estimated factors $\widehat\bF_t$ (up to a transformation). It shows that, in order to estimate the factors consistently, the signal to noise ratio of the factor model ($\lambda/\sigma$) must go to infinity, in order to have sufficient information on the signal part at each time point $t$.
Concerning the impact of $\lambda/\sigma$, it is not surprising that the stronger the factor strength is, the more useful information the observed process carries and the faster the estimated factors converge.
Moreover, \eqref{eq:prop:factor1} shows that the error rates for the sample auto-covariance of the estimated factors to the true sample auto-cross-moment is $o_{\P}(T^{-1/2})$ when $\lambda/\sigma \gg \sqrt{d_{\max}}$.
When $\lambda/\sigma\gg \sqrt{d_{\max}}+\sqrt{T}$, \eqref{eq:prop:factor2} is much smaller than the parametric rate $T^{-1/2}$. Hence, intuitively, this implies that it is a valid option to use the estimated factor processes as the true factor processes to model the dynamics of the factor under such situations. The results are expected to be the same as using the true factor process, without loss of efficiency.
The statistical rates in \eqref{eq:prop:factor1} and \eqref{eq:prop:factor2} lay a foundation for further modeling of the estimated factor processes with vast repository of linear and nonlinear options.

\begin{rmk}
Under the setting of \cite{wang2019}, $\lambda/\sigma\asymp d_1^{1/2-\delta_1/2}d_2^{1/2-\delta_2/2}$ for some $\delta_1>0, \delta_2>0$. The bound in Proposition \ref{prop:factor} becomes $$\left\| \widehat\bF_t - \bR_1 \bF_t \bR_2^\top \right\|_{\rm S}=O_{\P}(d_{\min}^{-1}d_1^{\delta_1/2}d_2^{\delta_2/2}T^{-1/2}+d_1^{-1/2+\delta_1/2}d_2^{-1/2+\delta_2/2}),\quad d_{\min}=\min\{d_1,d_2\}.$$
In comparison, the first term of the above bound is much sharper than the results in Theorem 3 in \cite{wang2019}, while the second term keeps the same. It also demonstrates the benefits of using iterative procedures for the first stage factor loading matrices estimation over using non-iterative procedures.
\end{rmk}

The following theorem shows the rate of convergence for alternating least square estimators of the second stage MAR modelling of the latent factor process, directly using $\hat{\bF}_t$ as observed, outlined in Section~\ref{sec:standard}. Due to the identifiability issue, we make the convention that $\|\bA_1\|_{\rm F}=1, \|\widehat \bA_1\|_{\rm F}=1.$

\begin{theorem}\label{thm:a}
Consider model \eqref{eq:dmfm1} and \eqref{eq:dmfm2}. Suppose Assumptions \ref{asmp:error} and \ref{asmp:mar-noise} hold. Assume $r_1,r_2$ are fixed. Also assume the causality condition $\rho(\bA_1) \rho(\bA_2) < 1$. {There exit rotation matrices $\bR_1, \bR_2$} 
that, for the iterative least square estimators of MAR(1), it holds
\begin{align}\label{thm:a:eq}
\left(\begin{matrix} \vec(\widehat \bA_1^{(\rm LSE)} - \bR_1\bA_1\bR_1^\top ) \\ \vec(\widehat \bA_2^{(\rm LSE)\top} - \bR_2^\top \bA_2^{\top}\bR_2 ) \end{matrix} \right) =O_{\P} \left(\frac{\sigma \sqrt{d_{\max}} }{\lambda\sqrt{T}} + \frac{\sigma^2}{\lambda^2} +\frac{1}{\sqrt{T}} \right).
\end{align}
\end{theorem}

The proof of the theorem is given in Appendix \ref{append:proof}.

The first two terms of \eqref{thm:a:eq} come from the estimation error of the latent factor process and the covariance matrices. Compared with \eqref{eq:prop:factor}, the errors are more precisely controlled in the second stage estimation of the MAR component.

In addition to the consistency of the MAR estimators, we also have the following asymptotic normality under a stronger signal strength condition.

\begin{theorem}\label{thm:b}
Consider model \eqref{eq:dmfm1} and \eqref{eq:dmfm2}. Suppose Assumptions \ref{asmp:error} and \ref{asmp:mar-noise} hold. Assume $r_1,r_2$ are fixed. Also assume the causality condition $\rho(\bA_1) \rho(\bA_2) < 1$. Let $\Sigma=\Cov(\vec(\bxi_t))$, $\balpha_1:=\vec(\bA_1)\in \R^{r_1^2}$, 
$\bgamma:=(\balpha_1^{\top},{\bf 0}^{\top})^{\top}$, and $d_{\max}=\max\{d_1,d_2\}$. Define $\bW_t^{\top}=[(\bA_2\bF_t^{\top}) \otimes \bI_{r_1}; \bI_{r_2} \otimes (\bA_1\bF_t)]$, $\bH_1:=\E(\bW_t \bW_t^{\top})+\bgamma\bgamma^{\top}$, and $\Xi_1:=\bH_1^{-1} \E (\bW_t \Sigma \bW_t^{\top}) \bH_1^{-1}$, where $\otimes$ is Kronecker product. If $\lambda/\sigma \gg T^{1/4}+ \sqrt{d_{\max}}$, then there exit rotation matrices $\bR_1, \bR_2$ such that the iterative least square estimators of MAR(1) satisfy,
\begin{align}
\sqrt{T} \left(\begin{matrix} \vec(\widehat \bA_1^{(\rm LSE)} - \bR_1\bA_1\bR_1^\top) \\ \vec(\widehat \bA_2^{(\rm LSE)\top} - \bR_2^\top \bA_2^{\top}\bR_2 ) \end{matrix} \right) \Rightarrow    N(0,\Xi_1).
\end{align}

\end{theorem}

The proof of the theorem is given in Appendix \ref{append:proof}.

\begin{rmk}
The central limit theorem for the least square estimators $\widehat\bA_1^{(\rm LSE)},\widehat\bA_2^{(\rm LSE)}$ holds only when the signal strength is sufficiently strong (i.e. when $\lambda/\sigma\gg \sqrt{d_{\max}}+ T^{1/4}$).
For weaker signal strengths, we are only able to establish consistency, as the estimation error of the factor loading matrices and the idiosyncratic noise $\bE_t$ will dominate the estimation error of MAR(1). Additionally, under such conditions, the estimation error of the sample covariance matrix of the factor processes in \eqref{eq:prop:factor2} will dominate the parametric rate. This complication makes the asymptotic distribution challenging to derive.
\end{rmk}

\begin{rmk}
Strong factor model in the literature \citep{lam2011,bai2003} requires that $\lambda/\sigma \asymp \sqrt{d_1 d_2}$, where $\lambda$ pools all the information (singular values) from the conventional loading matrices and model \eqref{eq:dmfm1} assumes orthogonal $\bU_1,\bU_2$. Thus, if $d_1d_2\gg \sqrt{T}$
asymptotic normality in Theorem \ref{thm:b} holds. This relationship between sample size and dimensions may seem counter-intuitive at first glance. We will justify the intuition as follows. When $d_1d_2$ is very large, we may view the estimated loading matrices $\widehat\bU_k$ to be the same as the true loading matrices $\bU_k$. Then, $\widehat\bF_t=\bF_t+\bU_1^{\top} \bE_t \bU_2/\sqrt{d_1d_2}$, by setting $\lambda=\sqrt{d_1d_2}$. The second term in $\widehat\bF_t$, i.e. $\bU_1^{\top} \bE_t \bU_2/\sqrt{d_1d_2}$, is still possible to contribute a dominating error term, larger than the parametric error rate $o_{\P}(T^{-1/2})$, unless $d_1d_2$ are sufficiently large.
Such phenomenon also appears in the factor model literature; see for example \cite{fan2013}.
\end{rmk}

\begin{rmk}
To estimate loading matrices $\bU_k$ and latent factor process $\bF_t$ consistently, we need $\sigma d_{\max}^{1/2}\lambda^{-1}T^{-1/2}+\sigma/\lambda=o(1)$ which allows finite sample $T$. However, based on Theorems \ref{thm:a} and \ref{thm:b}, $T\to \infty$ is required for consistent estimation of $\bA_1,\bA_2$.
\end{rmk}

Furthermore, we have the following convergence rates and central limit theorems for the Lag-2
estimators, $\widehat \bA_1^{(\rm L2E)}$ and $\widehat \bA_2^{(\rm L2E)}$
in Section~\ref{sec:lagged}.

\begin{theorem}\label{thm:laga}
Consider model \eqref{eq:dmfm1} and \eqref{eq:dmfm2}. Suppose Assumptions \ref{asmp:error} and \ref{asmp:mar-noise} hold. Assume $r_1,r_2$ are fixed. Also assume the causality condition $\rho(\bA_1) \rho(\bA_2) < 1$.
There exit rotation matrices $\bR_1, \bR_2$, it holds that for the Lag-2 estimator
of MAR(1),
\begin{align}
\left(\begin{matrix} \vec(\widehat \bA_1^{(\rm L2E)} - \bR_1\bA_1\bR_1^\top ) \\ \vec(\widehat \bA_2^{(\rm L2E)\top} - \bR_2^\top \bA_2^{\top}\bR_2 ) \end{matrix} \right) =O_{\P} \left(\frac{\sigma \sqrt{d_{\max}} }{\lambda\sqrt{T}} +\frac{\sigma^2 }{\lambda^2\sqrt{T}} + \frac{1}{\sqrt{T}} \right).
\end{align}
\end{theorem}

\begin{theorem}\label{thm:lagb}
Consider model \eqref{eq:dmfm1} and \eqref{eq:dmfm2}. Suppose Assumptions \ref{asmp:error} and \ref{asmp:mar-noise} hold. Assume $r_1,r_2$ are fixed. Also assume the causality condition $\rho(\bA_1) \rho(\bA_2) < 1$. Let $\Sigma=\Cov(\vec(\bxi_t))$, $\balpha_1:=\vec(\bA_1)\in \R^{r_1^2}$, 
$\bgamma:=(\balpha_1^{\top},{\bf 0}^{\top})^{\top}$ and $d_{\max}=\max\{d_1,d_2\}$.
Define $\bG_t=\bA_1\bF_t \bA_2$, $\bQ_t^{\top}=[(\bA_2\bG_t^{\top}) \otimes \bI_{r_1}; \bI_{r_2} \otimes (\bA_1\bF_t)]$, $\bH_2:=\E(\bQ_t \bQ_t^{\top})+\bgamma\bgamma^{\top}$, and $\Xi_2:=\bH_2^{-1} \E (\bQ_t \Sigma \bQ_t^{\top}) \bH_2^{-1}$, where $\otimes$ is Kronecker product.
If $\lambda/\sigma \gg \sqrt{d_{\max}}$, then there exit rotation matrices $\bR_1, \bR_2$ such that it holds for the Lag-2 estimator of MAR(1),
\begin{align}
\sqrt{T} \left(\begin{matrix} \vec(\widehat \bA_1^{(\rm L2E)} - \bR_1\bA_1\bR_1^\top) \\ \vec(\widehat \bA_2^{(\rm L2E)\top} - \bR_2^\top \bA_2^{\top}\bR_2 ) \end{matrix} \right) \Rightarrow    N(0,\Xi_2).
\end{align}
\end{theorem}

\begin{proposition}\label{prop:efficiency}
    Assume the conditions of Theorem~\ref{thm:lagb}. Assume the covariance matrix of $\bxi_t$ takes the form
$\Sigma=\Cov(\vec(\bxi_t))=\sigma_{\xi}^2\bI$.
It holds that $\Xi_1\preceq\Xi_2$, meaning $\Xi_2-\Xi_1$ is a positive semi-definite matrix.
\end{proposition}

\begin{rmk}
Compared to Theorem \ref{thm:a}, Theorem \ref{thm:laga} features an estimator with one less term in the rate and does not require $\lambda/\sigma\gg 1$. Similarly, compared to Theorem \ref{thm:b}, Theorem \ref{thm:lagb} does not require $\lambda/\sigma \gg T^{1/4}$. Hence when the SNR is not sufficiently high, one should use L2E.
On the other hand, when the SNR is high, we believe L2E is not as efficient as the LSE. Proposition \ref{prop:efficiency} shows that in the special case of $\Sigma=\sigma_{\xi}^2\bI$, we have $\Xi_1\preceq\Xi_2$, though it is difficult to show it in the more general case 
due to the complicated form of the asymptotically covariance matrices. Empirical evidence does support this conjecture.
{This is not surprising because it has been shown that the LSE is more efficient than L2E in the general vector time series setting.}
We reiterate that Theorem~\ref{thm:b} only holds under the condition $\lambda/\sigma\gg T^{1/4}$. If we consider a borderline case $\lambda/\sigma \asymp T^{1/4}$, it can be shown that $\sqrt{T}\,\vec(\hat\bA_i-\bA_i)$ converges to a normal distribution with nonzero mean. This asymptotic bias would only become larger if $\lambda/\sigma$ gets smaller.
\end{rmk}

\section{Numerical studies}

\subsection{Simulations settings}

To evaluate the performance of the proposed estimation and prediction procedures, we use the original model formulation \eqref{eq:dmfm1} and \eqref{eq:dmfm2} as the data generating mechanism.
For each $(d_i,r_i)$, the loading matrices $\bU_i$ is 
randomly generated as the first $r_i$ left singular vectors of a standard $d_i\times d_i$ Gaussian ensemble (with i.i.d. $N(0,1)$ entries).
The matrix $\bA_i$ is generated as $\bA_i=\bL_i\bD_i\bR_i^\top$, where $\bL_i$ and $\bR_i$ are $r_i\times r_i$ orthogonal matrices generated in the same way as $\bU_i$, and $\bD_i$ is a diagonal matrix whose diagonal entries are randomly generated from $\rm{Unif}[0.5,1.5]$. The matrices $\bA_1$ and $\bA_2$ are then rescaled so that $\|\bA_1\|_{\rm F} = 1$ and the spectral radius of $\bA_2\otimes\bA_1$ is $\rho$.
Throughout the simulations, $\bE_t$ is a $d_1 \times d_2$ standard Gaussian ensemble.
The innovations $\bxi_t$ in \eqref{eq:dmfm2} are generated as i.i.d. Gaussian with covariance matrix $\bSigma_\bxi=\Cov(\vec(\bxi_t))$. The $\bSigma_\bxi$ is generated through its eigenvalue decomposition, similar to the generation of $\bA_i$, but its $r_1r_2$ eigenvalues are equally spaced over $[1,10]$.
For the original DMFM formulation \eqref{eq:dmfm1} and \eqref{eq:dmfm2}, the SNR is defined as
\begin{align*}
\text{SNR}:= \sqrt{\frac{\E \|  \lambda\bF_{t}  \|^2_{\rm F}} {\E \|  \bU_1^\top\bE_{t}\bU_2  \|^2_{\rm F}}}.
\end{align*}
We will adjust the value of $\lambda$ to reach a specified level of SNR. In the simulation studies, we fix the factor dimensions to be $r_1=r_2=3$ and consider various configurations of $\{d_1, d_2, T, \rho, \lambda\}$. For each configuration, all the parameters related to the data generating mechanism, including $\bU_i$, $\bA_i$, $\lambda$, $\rho$ and $\bSigma_\xi$ are held fixed once they are generated. We will then simulate the DMFM model with 100 repetitions to inspect the empirical performance of the estimation in Section~\ref{sec:sim_est} and of the prediction in Section~\ref{sec:sim_pred}.

\subsection{Estimation performance}
\label{sec:sim_est}

The estimation of $\bU_i$ is adopted from \cite{chen2022factor} and \cite{han2020}, and is not the focus of this paper, so we only examine the the estimation of $\bA_i$ in this section.

To introduce the metric for the estimation performance, we note that the estimation of the loading $\bU_i$ is actually estimating the column space $\bU_i$. In other words, the $\hat\bU_i$ is trying to estimate $\bU_i$ after a orthogonal rotation. Although further identification conditions can be introduced to fully identify $\bU_1$ and $\bU_2$ up to a sign change, such an identification will only produce an equivalent estimation of $\bA_i$ and does not affect the prediction of $\bX_t$. Therefore, we shall skip the unnecessary additional identification. On the other hand, when $\bF_t$ are estimated as $\hat\bU_1^\top\bX_t\hat\bU_2$, the MAR model on $\bF_t$ can be written as
\begin{equation*}
    \hat\bU_1^\top\bU_1\bF_t\bU_2^\top\hat\bU_2 = [\hat\bU_1^\top\bU_1\bA_1(\hat\bU_1^\top\bU_1)^{-1}] [\hat\bU_1^\top\bU_1 \bF_{t-1} \bU_2^\top\hat\bU_2 ][(\bU_2^\top\hat\bU_2)^{-1}\bA_2^\top\bU_2^\top\hat\bU_2] + \hat\bU_1^\top\bU_1\bxi_t\bU_2^\top\hat\bU_2
\end{equation*}
Therefore, if an MAR model is fitted to $\hat\bF_t$, we are essentially estimating
\begin{equation*}
    \bA_{i0}:=\hat\bU_i^\top\bU_i\bA_i(\hat\bU_i^\top\bU_i)^{-1},\quad i=1,2.
\end{equation*}
In the simulation we will use the squared error (on log scale)
to evaluate the performance of different estimators:
\begin{equation}
\label{eq:log_se}
    \log\|\hat\bA_2\otimes\hat\bA_1 - \bA_{20}\otimes\bA_{10}\|_{\rm F}^2,
\end{equation}
where $\hat\bA_i$ is either the LSE or the L2E.

\begin{figure}[htbp]
\centering
\includegraphics[width=\textwidth]{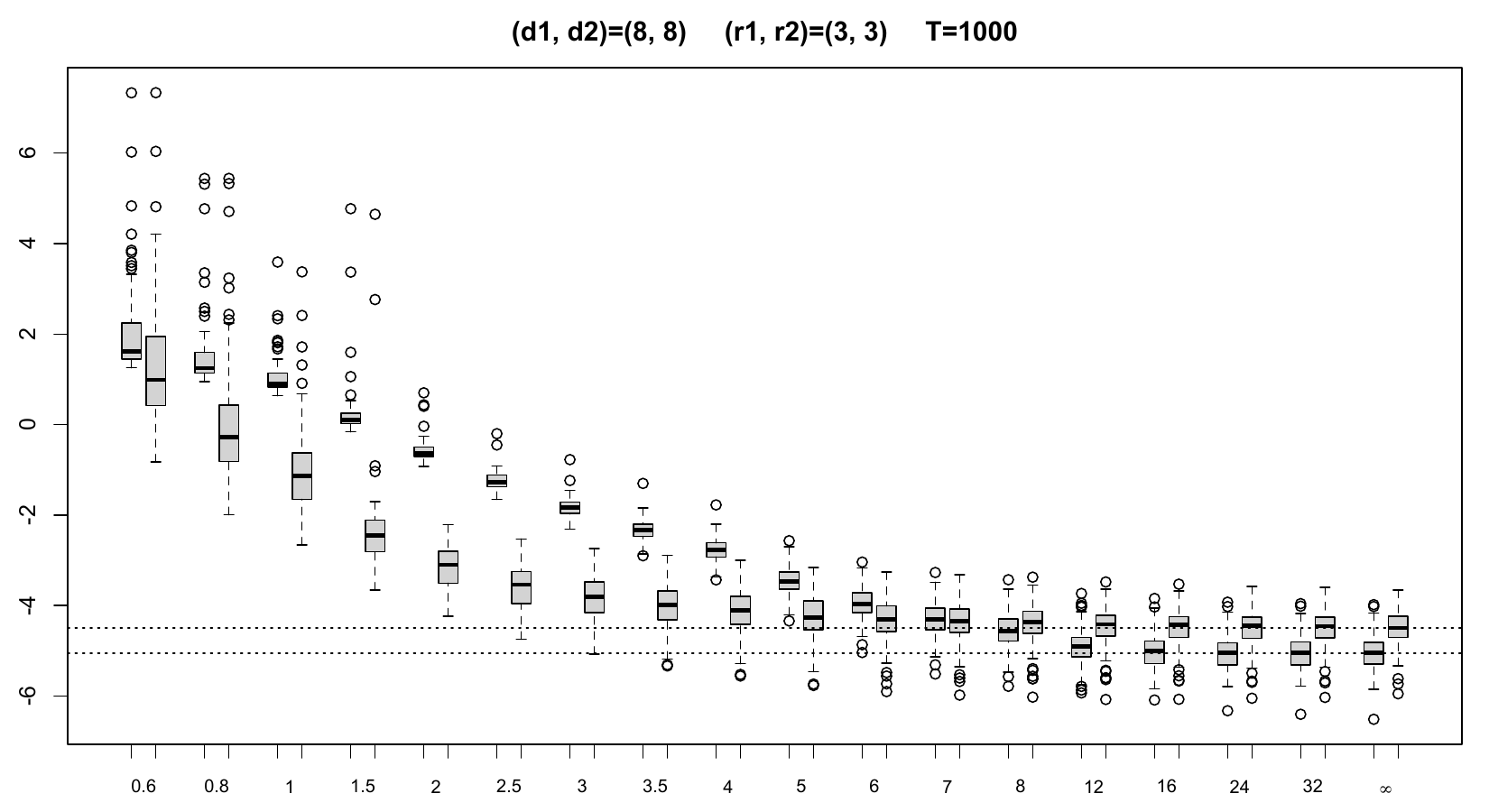}
\caption{Log squared errors of LSE and L2E. The SNR is labelled at the horizontal axis. Two dashed lines mark the median log squared error of the oracle LSE and L2E. \label{fig:sim_est_1}}
\end{figure}

In Figure~\ref{fig:sim_est_1}, we show the box plots of the errors over 100 repetitions \eqref{eq:log_se} for the case $d_1=d_2=8, r_1=r_2=3, \rho=.9$ and $T=1000$. The box plot is given against the SNR ranging from 0.6 to 128, as labelled at the horizontal axis. For each SNR, two box plots are shown, the one on the left for LSE, and the one on the right for L2E. The two horizontal lines show the median errors
(corresponding to the two box plots at the very right side)
for the oracle case where the true $\bF_t$ are used to estimate $\bA_i$. This is equivalent to an infinity SNR so there is no estimation error in $\hat\bF_t$. We label it as $\infty$ in the plot.

There are a number of observations that can be made from Figure~\ref{fig:sim_est_1}. First, we see the general pattern that when the SNR is low, L2E performs better than the LSE, and when the SNR reaches 8, LSE starts to outperform L2E. This is consistent with the discussion made in Section~\ref{sec:lagged} and the Proposition~\ref{prop:efficiency}. Second, L2E can be significantly better than LSE, especially when the SNR is between 0.8 and 5. We note that the errors are plotted in the log scale, so a unit difference in Figure~\ref{fig:sim_est_1} means the squared error is actually reduced by a factor of $1/e$. Last but not least, we see that as the SNR increases, eventually both the LSE and the L2E are getting closer to the oracle estimators when there is no estimation error. It is clear that L2E is less efficient in the oracle case, as Proposition~\ref{prop:efficiency} indicates. We would like to further point out that the L2E starts to behave almost same as its oracle version when the SNR reaches 8, while the LSE requires a SNR as large as 16 to behave like the oracle LSE. This again confirms the fact the L2E is less susceptible to the estimation errors in $\hat\bF_t$.

Similar plots for different simulation settings are give in the appendix.

\subsection{Prediction Performance}
\label{sec:sim_pred}

In this section, we compare the performance of five prediction methods based on the DMFM. Throughout this section we consider one-step ahead prediction with origin $T$.
\begin{enumerate}
    \item [(a)] LSE. Plug-in prediction \eqref{eq:pred_plug} using LSE.
    \item [(b)] L2E. Kalman filter based prediction \eqref{eq:pred_kf} using L2E.
    \item [(c)] V.LSE. Fitting a VAR(1) $\vec(\bF_t) = \bPhi\vec(\bF_{t-1})+\vec(\bxi_t)$ model to $\vec(\hat\bF_t)$ using least squares, and predict $\bX_{T+1}$ by $\hat\bX_T(1)=\hat\bU_1\mat_1\left[\hat\bPhi^{\rm (LSE)}\vec(\hat\bF_T)\right]\hat\bU_2^\top$.
    \item [(d)] V.L2E. When fitting the VAR to $\hat\bF_t$, estimate $\bPhi$ by $\bPhi^{\rm (L2E)}=\hat\bGamma_2\hat\bGamma_1^{-1}$, and predict $\bX_{T+1}$ by $\hat\bX_T(1)=\hat\bU_1\mat_1\left[\hat\bPhi^{\rm (L2E)}\vec(\hat\bF_T)\right]\hat\bU_2^\top$.
    \item [(e)] L2E+. Predict $\bX_{T+1}$ by \eqref{eq:pred_kf} only if the smallest eigenvalue of $\hat\bSigma_\bzeta$ is positive, and by \eqref{eq:pred_plug} otherwise. This is the method we recommend at the end of Section~\ref{sec:pred}.
\end{enumerate}
We compare the methods listed above through the prediction mean squared error
\begin{equation}
\label{eq:pse}
    \hbox{PSE}:=(d_1d_2)^{-1}\|\hat\bX_T(1)-\bU_1\bA_1\bF_T\bA_2^\top\bU_2^\top\|^2/(\hbox{SNR})^2.
\end{equation}
Note that in the simulation for different SNR, we fix the generating mechanism of $\bE_t$, and change the value of $\lambda$ in \eqref{eq:dmfm1} in order to reach different levels of SNR. Consequently, there is a factor $(\hbox{SNR})^{-2}$ in the definition of PSE in \eqref{eq:pse} so that the PSE at different SNR levels are comparable.

\begin{figure}[htbp!]
\centering
\includegraphics[width=\textwidth]{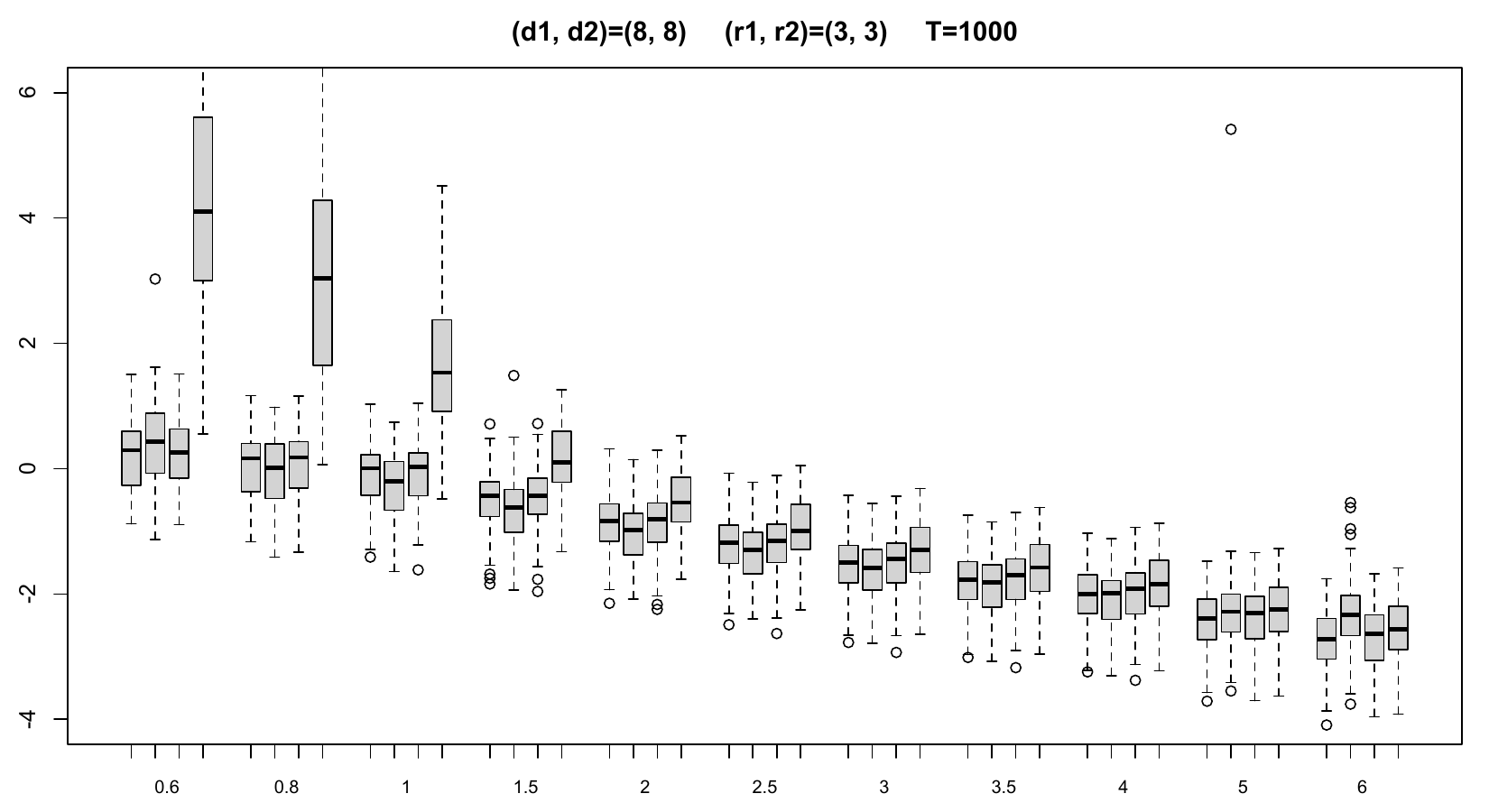}
\caption{Log squared prediction errors of LSE, L2E, V.LSE and V.L2E. The SNR is labelled at the horizontal axis.  \label{fig:sim_pred_1}}
\end{figure}

\begin{figure}[htbp!]
\centering
\includegraphics[width=.5\textwidth]{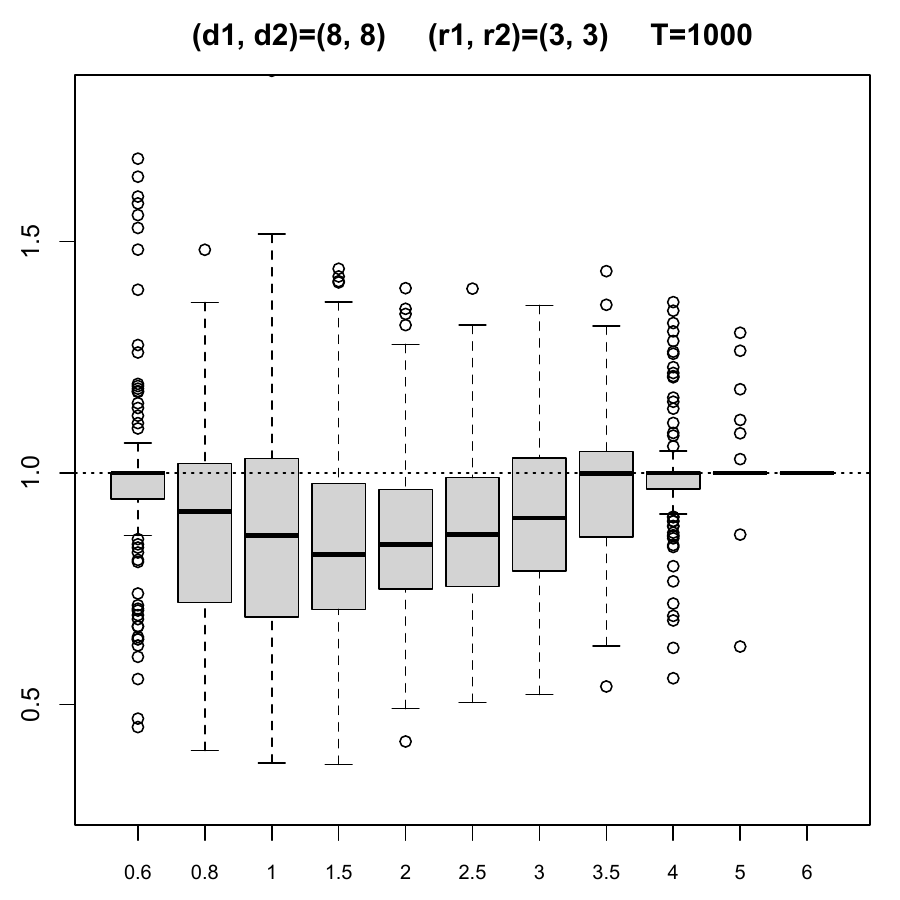}\includegraphics[width=.5\textwidth]{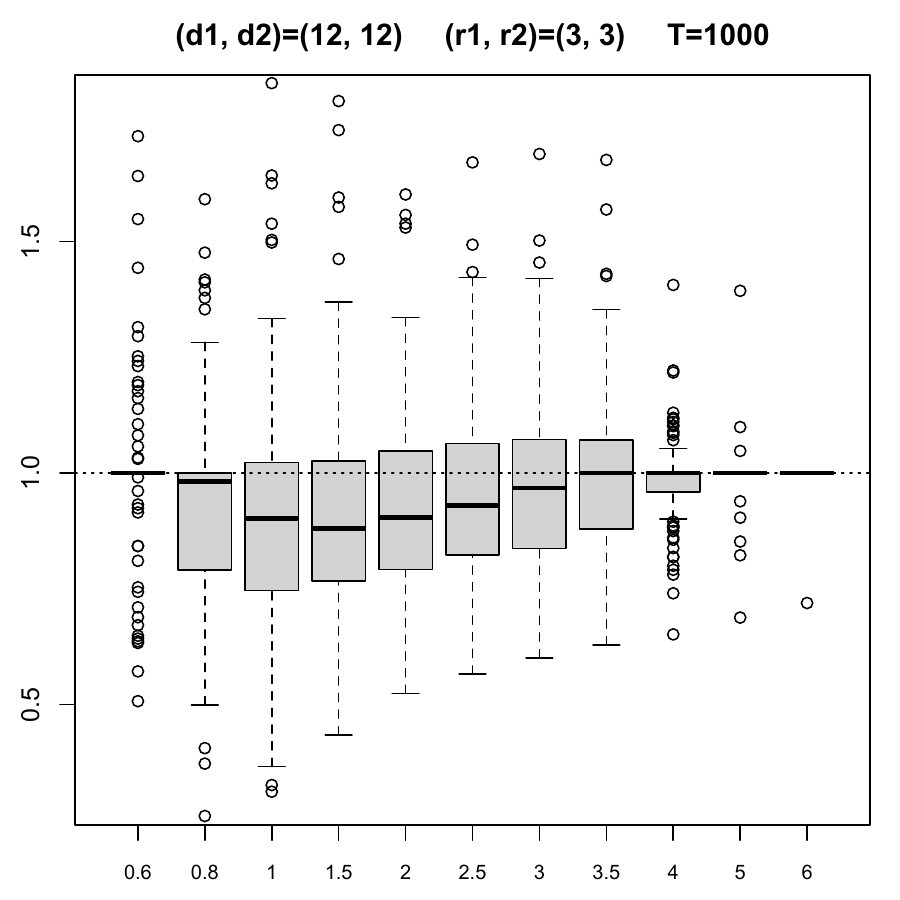}
\caption{Ratios of the squared prediction errors of L2E+ over LSE. The SNR is labelled at the horizontal axis.  \label{fig:sim_pred_2}}
\end{figure}

In Figure~\ref{fig:sim_pred_1} we compare the first 4 methods using the box plots of $\log(PSE)$ over 100 repetitions. For each SNR, 4 box plots are for LSE, L2E, V.LSE, V.L2E from left to right. We see that among LSE, V.LSE and V.L2E, the LSE prediction is consistently better than or comparable to others. Comparing LSE and L2E, we see L2E enjoys a better performance when the SNR is between 0.8 and 4. When the SNR is larger, the LSE leads to the best predictions. We also compare the LSE and L2E+ in Figure~\ref{fig:sim_pred_2}, by plotting the ratios of the L2E+ PSE over the LSE PSE. These ratios are in general more likely to be below one, confirming the advantage of the recommended L2E+ prediction method, which adaptively combines the strength of LSE and L2E. When the SNR is as large as 5 or 6, L2E+ starts to coincide with LSE.

\section{Real example: New York City Yellow Cab Taxi Trips}
In this section, we apply the dynamic matrix factor model to analyze 
the New York City yellow cab taxi traffic data. This data is maintained by the Taxi \& Limousine Commission of New York City and can be downloaded from https://www1.nyc.gov/site/tlc/about/tlc-trip-record-data.page. The portion of the dataset we use comprises more than 1 billion trip records spanning from January 1, 2009, to December 31, 2019. Most of the trips occur within Manhattan Island. Each trip record includes the time and location of pickup and drop-off, as well as payment information.

\begin{figure}[htbp!]
\begin{center}
  \caption{Manhattan taxi zone map}\label{fig:manhattan taxi map}
\includegraphics[scale=0.15]{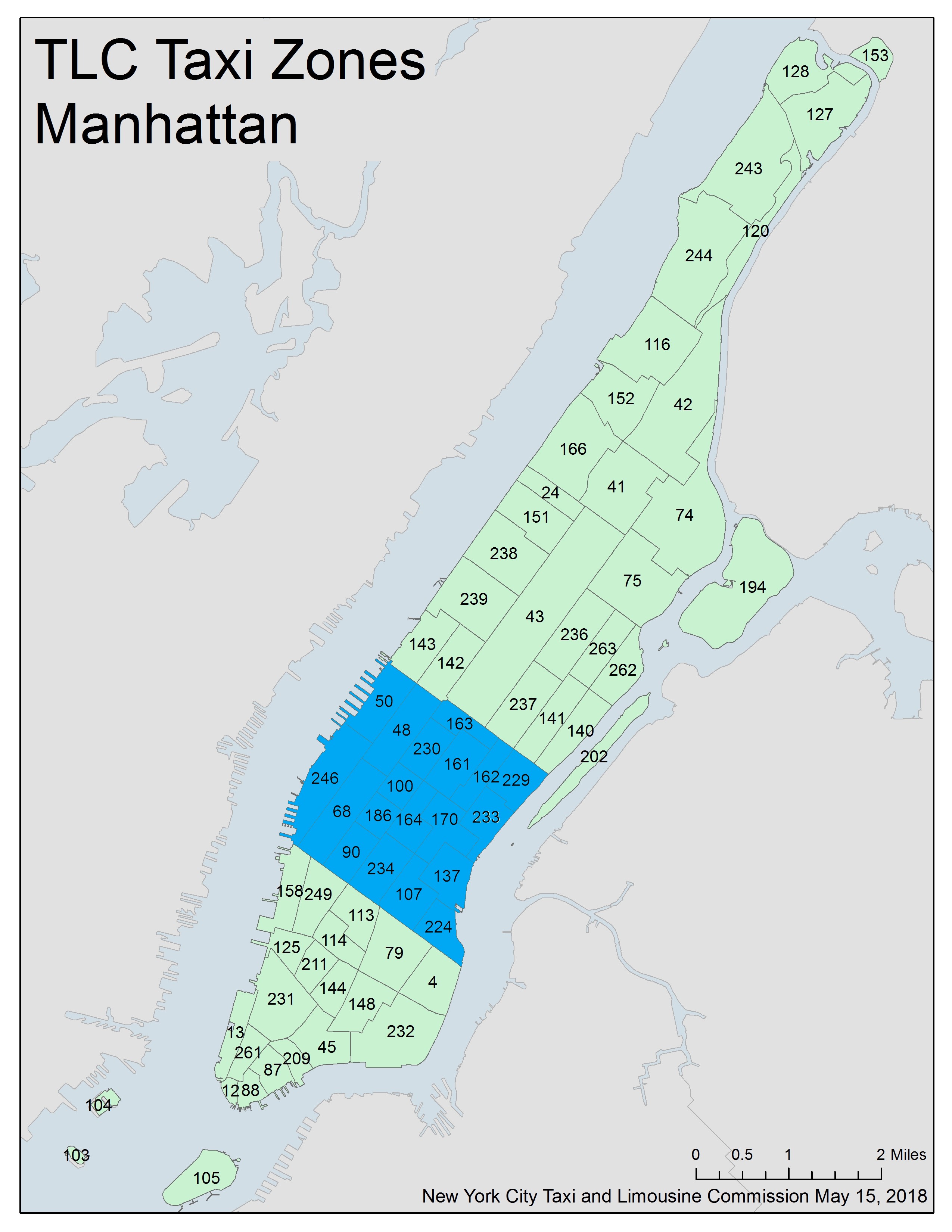}
\end{center}
\end{figure}

Similar to that in \cite{chen2022factor}, we obtain a
$19\times 19\times 750$ matrix time series, with 19 of 69 predefined zones in Manhattan (see the blue area in the map in Figure \ref{fig:manhattan taxi map} for more details) and $T=750$ business days in years 2015-2017.  The zones selected are mainly the midtown areas of Manhattan, which include Penn Station and numerous busy business districts. Each entry $X_{ijt}$ is the aggregated number of ride from zone $i$ to zone $j$ in the hours between 7am to 10am (morning rush hours) on day $t$.

\begin{figure}[ht]
\begin{center}
  \caption{Number of trips during morning rush hours in selected regions} \label{fig:daily total trips}
  \includegraphics[scale=0.4]{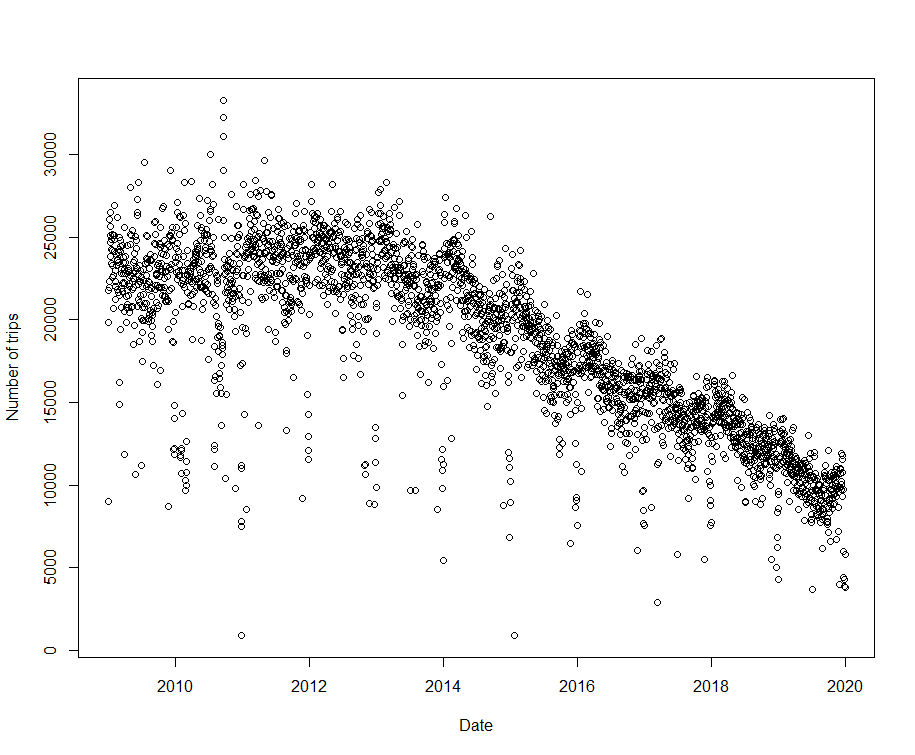}
\end{center}

\end{figure}
\begin{figure}[htbp!]
  \caption{2015-2017, 7-10am morning rush hours detrended data}  \label{fig: time series plot}
  \includegraphics[scale=0.6]{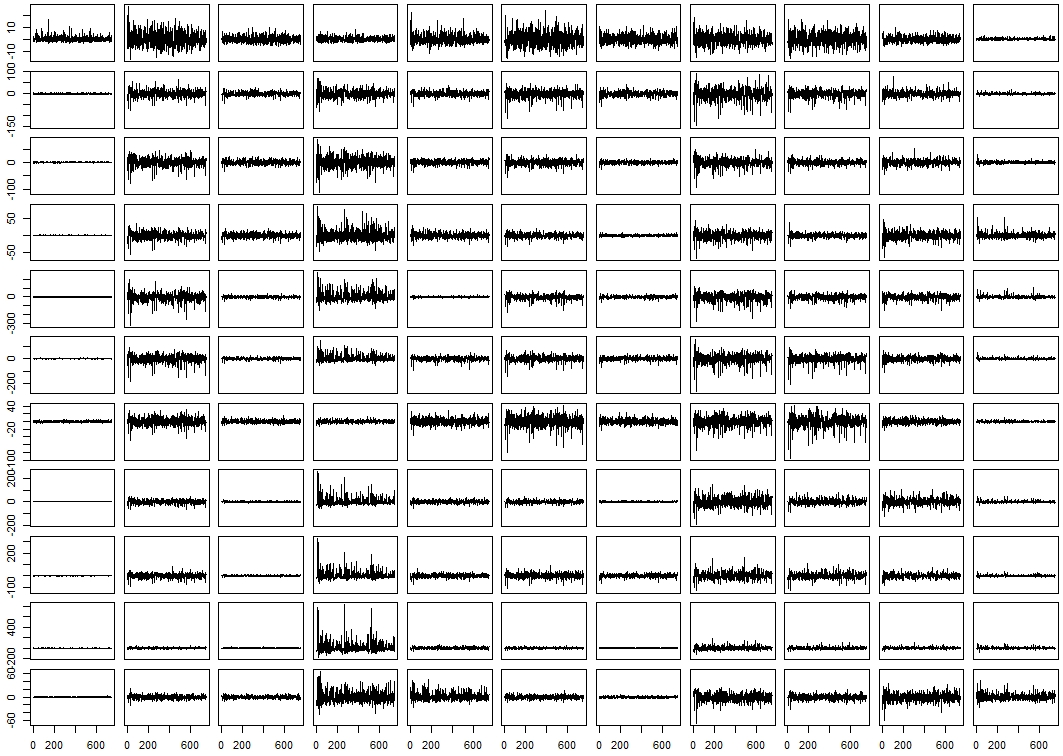}
\end{figure}

Figure \ref{fig:daily total trips} plots the daily sum of trips across all 19 selected regions in Manhattan during 7am to 10am, revealing a decreasing trend starting from 2014. This decline can be attributed to the emergence of ride-hailing services. 
Additionally, seasonal trends within each year are visible, likely due to temperature fluctuations. These two factors contribute to the non-stationarity of the time series, necessitating the removal of trends using exponential smoothing and the subsequent building of models on the residual data. Specifically, using smoothing parameter $\alpha=0.1$, we obtain the exponential smoothing trend  $\hat{\bB}_{t} = \alpha \bY_t + (1-\alpha) \hat{\bB}_{t-1}$ where $\bY_t$ is observed matrix time series. Let $\bX_t=\bY_t-\bB_t$ as the detrended series. Figure \ref{fig: time series plot} shows the detrended series among 11 zones. The $i$-th rows shows the pick-up volume from $i$-th zone to one of the other zones, and the $j$-th column show the drop-off volume in $j$-th zone coming from one of the other zones. 
Each individual time series is mostly stationary after removing the trends. We can see that some zones, such as Zone 1 (the first zone from the 19 selected) have large pickup volumes, but very small drop-off volumes. These zones are mostly residential areas, with people going to work 
during morning rush hours. Zones such as 4 and 8, however, have large drop-off volumes but small pickup volumes, are commercial districts.


Dynamic matrix factor model and several other models are
estimated and their out-sample rolling forecasting performance
are comparied.
The rolling period is the entire year 2017 period.
For each day in year 2017, we use all the past observations to fit the model, and make a prediction of the current day and compare with the true trip counts on that day. We report
\[
\mbox{RMSE} =  \sqrt{(250\times d_1\times d_2)^{-1}\sum\limits_{t=499}^{749}\|\hat{\bX}_{t}(1) - \bX_{t+1}\|_{\rm F}^2}.
\]

\begin{table}[!htbp]
\caption{Compare Rolling Forecasts Among Different Models}
\centering
\begin{tabular}{|c|c||c|c||c|c|c|}
\hline
\multicolumn{2}{|c||}{Matrix Factor Models} & \multicolumn{2}{c||}{Vector Factor Models} & \multicolumn{2}{c|}{Autoregressive Models} \\ \hline
Method & RMSE  & Method       & RMSE  & Method         & RMSE  \\ \hline
\bf MFMLSE & \bf 12.68 & / & / & MAR & 12.71 \\ \hline
MFML2E & 13.16 & / & / & / & / \\ \hline
/ & / & / & / & MARR & 12.72 \\ \hline
MFV & 12.75 & VFV & 12.71 & VAR & 19.05 \\ \hline
MFi & 12.81 & VFi   & 12.74  & iAR & 12.86 \\ \hline
MFrw   & 14.69 & VFrw & 14.90 & Xrw   & 16.12 \\ \hline
MFmean & 13.34 & VFmean & 13.33 & Xmean & 13.35 \\ \hline
\end{tabular}\label{table: taxi results}
\end{table}



Here is a list of models we compared.

\noindent
(a) \textbf{Matrix factor model methods:} The first step is to estimate a matrix factor model with factor rank $(5,5)$ and obtain the estimated factor process $\hat{\bF}_t$. In the second step, various models are used to fit $\hat{\bF}_t$ for prediction, including a MAR model with LSE (MFMLSE),
a VAR model (MFV), individual independent AR model for each factor element (MFi), a random walk model for the factors (MFrw), and a constant mean model for the factors using the estimated mean (MFmean).



\noindent
(b) \textbf{Vector factor model methods:} The first step is to estimate a vector factor model with factor rank $25$ using the stacked observations $\vec(\bX_t)$ and obtain the estimated factor process $\hat{\boldsymbol{f}}_t$. In the second step, various models are used to fit $\hat{\boldsymbol{f}}_t$ for prediction, including a VAR model (VFV), individual independent AR model for each factor element (VFi), a random walk model for the factors (VFrw), and a constant model for the factors using the estimated mean (VFmean).

\noindent
(c) \textbf{Autoregressive model on the observed $\bX_t$:} These models do not apply a factor model, but instead directly apply MAR(MAR) of \cite{chen2021autoregressive}, reduced rank MAR (MARR) of \cite{han2023rr}, VAR using the stacked $\vec(\bX_t)$ (VAR), individual AR (iAR), random walk (Xrw), and constant time series using the mean estimator (Xmean).
In Table \ref{table: taxi results}, we can see that the dynamic matrix factor model has the best prediction performance among all models. The individual AR models (MFi, VFi and iAR) are surprisingly better than some of the more complicated models.
VAR performs the worst, as it tends to overfit with a $19^2\times 19^2$ coefficient matrix. This data set appears to have a large SNR, so the plug-in prediction \eqref{eq:pred_plug} based on the DMF performs the best. We also remark that the proposed rule at the end of Section~\ref{sec:pred} always choose to use the plug-in prediction over the Kalman-filter-based prediction \eqref{eq:pred_kf}.

\section{Conclusion}

In this paper, we present a dynamic matrix factor model as an extension of the matrix factor model. The model incorporates the dynamics of latent factors by using a matrix autoregressive model, making it more suitable for analyzing the dynamics of low-dimensional latent factors and making predictions. The dynamic matrix factor model is similar in form to the reduced rank matrix autoregressive model, with the main difference being the lower rank error. The model is estimated in two steps, starting with estimating the factor model and then incorporating the matrix autoregressive dynamics. To improve the estimation accuracy, we propose a lag-2 Yule-Walker estimator and use Kalman Filter to reduce the influence of errors from matrix factor estimation when the signal-to-noise ratio is low. This allows us to obtain more accurate estimates of the underlying factors and their dynamics, even in the presence of noise and other sources of measurement error. The prediction is
done through the prediction of the latent factors.
Our numerical analysis demonstrates that the dynamic matrix factor model performs well in terms of estimation and prediction performance 
by capturing the underlying dynamics of the latent factors.
We also apply the dynamic matrix factor model to NYC yellow cab taxi data to demonstrate its advantages.
Our results suggest that the dynamic matrix factor model is a powerful tool for analyzing the dynamics of matrix-valued time series data and making accurate predictions over time.


\bibliographystyle{apalike} 
\bibliography{dmfm}

\newpage
\appendix

\section{Proofs of Main Theorems} \label{append:proof}

\begin{proof}[\bf Proof of Proposition \ref{prop1:lag-1 yule-walker}]
By definition, we have
\begin{align*}
      \hat{\bGamma}_1 &= \frac{1}{T-1}\sum\limits_{t=2}^T \vec(\bF_t) \vec({\bF}_{t-1})^{\top} \\ &= \frac{1}{T-1}\sum\limits_{t=2}^T (\bA_2 \otimes \bA_1) \vec({\bF}_{t-1}) \vec({\bF}_{t-1})^{\top} + \frac{1}{T-1}\sum\limits_{t=2}^T \vec(\bxi_t) \vec({\bF}_{t-1})^{\top} \\ &=  (\bA_2 \otimes \bA_1) \hat{\bGamma}_0 + \frac{1}{T-1}\sum\limits_{t=2}^T \vec(\bxi_t) \vec({\bF}_{t-1})^{\top},
\end{align*}
thus
\begin{align*}
\hat{\bGamma}_1 \hat{\bGamma}_0^{-1} - \bA_2 \otimes \bA_1 = \frac{1}{T-1}\sum\limits_{t=2}^T \vec(\bxi_t) \vec({\bF}_{t-1})^{\top} \hat{\bGamma}_0^{-1}.
\end{align*}

Let $\mathcal{F}_t$ be the filtration of $\bF_{t},\cdots, \bF_1$, then the series $\Big\{\frac{1}{T-1}\sum\limits_{t=2}^T \vec(\bxi_t) \vec({\bF}_{t-1})^{\top} \hat{\bGamma}_0^{-1}\Big\}_{T\geq 2}$ is a martingale since $\bxi_t$ and $\vec({\bF}_{t-1})$ are uncorrelated. Furthermore, when the innovation series $\bxi_t$ are i.i.d., the variance of the series is
\begin{align*}
        {\rm Var} & \big[ \frac{1}{T-1}\sum\limits_{t=2}^T \vec(\bxi_{t}) \vec({\bF}_{t-1})^{\top} \hat{\bGamma}_0^{-1} | \mathcal{F}_{t-1} \big] \\
        &= \frac{1}{(T-1)^2}\E \big[\hat{\bGamma}_0^{-1} \sum\limits_{t=2}^T \vec({\bF}_{t-1}) \vec(\bxi_{t})^\top \vec(\bxi_{t}) \vec({\bF}_{t-1})^\top \hat{\bGamma}_0^{-1} | \mathcal{F}_{t-1} \big]  \\
        &= \frac{1}{(T-1)^2} \bGamma_0^{-1} (T-1)\sigma^2 \bGamma_0 \bGamma_0^{-1} = \frac{1}{T-1}\sigma^2 \bGamma_0^{-1}.
\end{align*}
Here $\mathbb{E} \hat{\bGamma}_0 = \bGamma_{0}$ by Ergodic Theorem and the condition that $\bF_t$ and $\bxi_{t}$ are uncorrelated. By Martingale Central Limit Theorem \citep{HALL198051}, we have
\begin{align*}
\frac{1}{\sqrt{T-1}} \sum\limits_{t=2}^T \vec(\bxi_t) \vec({\bF}_{t-1})^{\top} \hat{\bGamma}_0^{-1} \Rightarrow N(0,\sigma^2 \bGamma_0^{-1}).
\end{align*}
\end{proof}

\begin{proof}[\bf Proof of Proposition \ref{prop2:lag-2 yule-walker}]
By the definition of lag-2 sample covariance matrix $\hat{\bGamma}_2$, We have
\begin{align*}
\hat{\bGamma}_2 &=
\frac{1}{T-2}\sum\limits_{t=3}^T \vec({\bF}_t) \vec(\bF_{t-2})^{\top} \\ &= \frac{1}{T-2}(\bA_2 \otimes \bA_1) \sum\limits_{t=3}^T \vec({\bF}_{t-1}) \vec({\bF}_{t-2})^{\top} + \frac{1}{T-2}\sum\limits_{t=3}^T \vec(\bxi_t) \vec({\bF}_{t-2})^{\top} \\ &= (\bA_2 \otimes \bA_1) \hat{\bGamma}_1 + \frac{1}{T-2}\sum\limits_{t=3}^T \vec(\bxi_t) \vec({\bF}_{t-2})^{\top}.
\end{align*}

Similarly, we focus on
\begin{align*}
 \bA_2 \otimes \bA_1 - \hat{\bGamma}_2 \hat{\bGamma}_1^{-1} = \frac{1}{T-2}\sum\limits_{t=3}^T \vec(\bxi_t) \vec({\bF}_{t-2})^{\top} \hat{\bGamma}_1^{-1}.
\end{align*}
Let $\mathcal{F}_t$ be the filtration of $\bF_{t},\cdots, \bF_1$, then the series $\Big\{\frac{1}{T-2}\sum\limits_{t=3}^T \vec(\bxi_t) \vec({\bF}_{t-2})^{\top} \hat{\bGamma}_1^{-1}\Big\}_{T\geq 2}$ is a martingale since $\bxi_t$ and $\vec({\bF}_{t-2})$ are uncorrelated. Furthermore, when the innovation series $\bxi_t$ are i.i.d., the variance of the series is
\begin{align*}
 {\rm Var} &\big[\frac{1}{T-2}\sum\limits_{t=3}^T \vec(\bxi_{t}) \vec({\bF}_{t-2})^{\top} \hat{\bGamma}_1^{-1} | \mathcal{F}_{t-2} \big] \\
 &=  \frac{1}{(T-2)^2}\E \big[ \hat{\bGamma}_1^{-\top} \sum\limits_{t=3}^T \vec({\bF}_{t-2}) \vec(\bxi_{t})^\top \vec(\bxi_{t}) \vec({\bF}_{t-2})^\top \hat{\bGamma}_1^{-1} | \mathcal{F}_{t-2} \big] \\ 
&=\frac{1}{T-2}\sigma^2 \bGamma_1^{-\top} \bGamma_0 \bGamma_1^{-1}.
\end{align*}
Here $\mathbb{E} \hat{\bGamma}_i = \bGamma_{i}$  by Ergodic Theorem and the condition that $\bF_t$ and $\bxi_{t+h}$ are uncorrelated, for $i=0,1$ and $h\in \mathbb{Z}$. By Martingale Central Limit Theorem \citep{HALL198051}, we have
\begin{align*}
    \frac{1}{\sqrt{T-2}} \sum\limits_{t=3}^T \vec(\bxi_t) \vec({\bF}_{t-2})^{\top} \hat{\bGamma}_1^{-1} \Rightarrow N(0,\sigma^2 \bGamma_1^{-\top} \bGamma_0 \bGamma_1^{-1}).
\end{align*}
\end{proof}

\begin{lemma}\label{lemma-error}
Suppose Assumptions \ref{asmp:error} and \ref{asmp:mar-noise} holds. Then
\begin{align}
&\left\|\sum_{t=2}^T \widehat \bU_1^{\top} \bE_t \widehat \bU_2 \bA_2\bF_{t-1}^{\top} \right\|_{\rm S}=O_{\P}(\sigma\sqrt{ Td_{\max}}),   \label{eq:lemma-error-1} \\
&\left\|\sum_{t=2}^T \widehat \bU_2^{\top} \bE_t^{\top} \widehat \bU_1 \bA_1\bF_{t-1} \right\|_{\rm S}=O_{\P}(\sigma\sqrt{ Td_{\max}}),   \label{eq:lemma-error-2} \\
&\sum_{t=2}^T\vec(\widehat\bU_1^{\top} \bE_t \widehat\bU_2)^{\top} \vec(\bF_{t-1})=O_{\P}(\sigma\sqrt{ Td_{\max}}), \label{eq:lemma-error-3}  \\
&\sum_{t=2}^T\vec(\widehat\bU_1^{\top} \bE_t \widehat\bU_2)^{\top} \vec(\bxi_{t})=O_{\P}(\sigma\sqrt{ Td_{\max}}), \label{eq:lemma-error-4}  \\
&\sum_{t=2}^T\vec(\widehat\bU_1^{\top} \bE_t \widehat\bU_2)^{\top} \vec(\widehat\bU_1^{\top} \bE_{t-1} \widehat\bU_2)^{\top}=O_{\P}(\sigma\sqrt{ Td_{\max}}). \label{eq:lemma-error-5}
\end{align}

\end{lemma}
\begin{proof}
Let $\widebar\E=\E(\cdot |\bF_{1},...,\bF_{T})$ and $\widebar\P=\P(\cdot |\bF_{1},...,\bF_{T})$.
We prove \eqref{eq:lemma-error-1}. By Lemma G.1(ii) in \cite{han2020}, there exist $U_1^{(\ell)}\in \R^{d_1\times r_1}, U_2^{(\ell')}\in \R^{d_2\times r_2}$, $1\le \ell\le N_{d_1r_1,1/8}:=17^{d_1r_1}, 1\le \ell'\le N_{d_2r_2,1/8}:=17^{d_2 r_2}$, such that $\|U_1^{(\ell)} \|_{\rm S}\le 1, \|U_2^{(\ell')} \|_{\rm S}\le 1$, and
\begin{align*}
\left\|\sum_{t=2}^T \widehat \bU_1^{\top} \bE_t \widehat \bU_2 \bA_2\bF_{t-1}^{\top} \right\|_{\rm S} &\le \sup_{\|\widetilde U_1\|_{\rm S}\le1,\|\widetilde U_2\|_{\rm S}\le 1 }  \left\|\sum_{t=2}^T \widetilde U_1^{\top} \bE_t \widetilde U_2 \bA_2\bF_{t-1}^{\top} \right\|_{\rm S}  \\
&\le 2\max_{\ell\le N_{d_1r_1,1/8}, \ell'\le N_{d_2 r_2,1/8}} \left\|\sum_{t=2}^T U_1^{(\ell)\top} \bE_t U_2^{(\ell')} \bA_2\bF_{t-1}^{\top} \right\|_{\rm S}.
\end{align*}
Elementary calculation shows that, conditioning on $(\bF_{1},...,\bF_{T-1})$, $\|\sum_{t=2}^T \widehat \bU_1^{\top} \bE_t \widehat \bU_2 \bA_2\bF_{t-1}^{\top} \|_{\rm S}$ is a $\sigma \|\sum_{t=2}^T \bF_{t-1} \bA_2^{\top} \bA_2 \bF_{t-1}^{\top} \|_{\rm S}^{1/2}$ Lipschitz function of $(\bE_2,...\bE_{T})$. Cauchy–Schwarz inequality and Sudakov-Fernique inequality imply that
\begin{align*}
\widebar\E \left\|\sum_{t=2}^T U_1^{(\ell)\top} \bE_t U_2^{(\ell')} \bA_2\bF_{t-1}^{\top} \right\|_{\rm S} \le 2\sqrt{2}\sigma\sqrt{r_1} \left\|\sum_{t=2}^T \bF_{t-1} \bA_2^{\top} \bA_2 \bF_{t-1}^{\top} \right\|_{\rm S}^{1/2}.
\end{align*}
By Gaussian concentration inequalities for Lipschitz functions \citep{borell1975brunn}
\begin{align*}
&\widebar\P\left( \left\|\sum_{t=2}^T U_1^{(\ell)\top} \bE_t U_2^{(\ell')} \bA_2\bF_{t-1}^{\top} \right\|_{\rm S} - 2\sqrt{2}\sigma\sqrt{r_{1}} \left\|\sum_{t=2}^T \bF_{t-1} \bA_2^{\top} \bA_2 \bF_{t-1}^{\top} \right\|_{\rm S}^{1/2} \ge \sigma\left\|\sum_{t=2}^T \bF_{t-1} \bA_2^{\top} \bA_2 \bF_{t-1}^{\top} \right\|_{\rm S}^{1/2} x\right)    \\
&\le e^{-\frac{x^2}{2}}.
\end{align*}
Hence,
\begin{align*}
&\widebar\P\left( \left\|\sum_{t=2}^T \widehat \bU_1^{\top} \bE_t \widehat \bU_2 \bA_2\bF_{t-1}^{\top} \right\|_{\rm S} - 2\sqrt{2}\sigma\sqrt{r_{1}} \left\|\sum_{t=2}^T \bF_{t-1} \bA_2^{\top} \bA_2 \bF_{t-1}^{\top} \right\|_{\rm S}^{1/2} \ge \sigma\left\|\sum_{t=2}^T \bF_{t-1} \bA_2^{\top} \bA_2 \bF_{t-1}^{\top} \right\|_{\rm S}^{1/2} x\right)  \\
&\le N_{d_1r_1,1/8}N_{d_2r_2,1/8} e^{-\frac{x^2}{2}}= 17^{d_1r_1+d_2r_2} e^{-\frac{x^2}{2}}.
\end{align*}
By taking $x\asymp \sqrt{d_1r_1}+\sqrt{d_2r_2}$, we can show
\begin{align*}
&\left\|\sum_{t=2}^T \widehat \bU_1^{\top} \bE_t \widehat \bU_2 \bA_2\bF_{t-1}^{\top} \right\|_{\rm S}=O_{\P}(\sigma\sqrt{Td_{\max}}).
\end{align*}
Employing similar arguments, we can prove \eqref{eq:lemma-error-2}, \eqref{eq:lemma-error-3}, \eqref{eq:lemma-error-4} and \eqref{eq:lemma-error-5}.
\end{proof}

\begin{lemma}\label{lemma-gradient}
Suppose Assumptions \ref{asmp:error} and \ref{asmp:mar-noise} holds. Let $\Phi=\bA_2\otimes \bA_1$, $d_{\max}=\max\{d_1,d_2\}$. For any sequence $\{\eta_T\}$ such that
\begin{align}\label{eq:lemma-gradient1}
\frac{\eta_T}{\sigma\sqrt{d_{\max}}/(\lambda\sqrt{T})+\sigma^2/\lambda^2+1/\sqrt{T}} \to \infty,
\end{align}
then
\begin{align}\label{eq:lemma-gradient2}
\P\left(\inf_{\|\widebar\Phi-\Phi\|_{\rm F}\ge \eta_T} \sum_{t=2}^T \left\| \vec(\widehat \bF_t) -\widebar\Phi \vec(\widehat \bF_{t-1}) \right\|_2^2 \le \sum_{t=2}^T \left\| \vec(\widehat \bF_t) -\Phi \vec(\widehat \bF_{t-1}) \right\|_2^2 \right)\to 0    .
\end{align}
\end{lemma}

\begin{proof}
Let $\widehat \bF_t=\widehat \bU_1^{\top} \bX_t \widehat \bU_2/\lambda$. Without loss of generality, assume $\bR_1=I_{r_1}$ and $\bR_2=I_{r_2}$. Then, write
\begin{align}
\vec(\widehat \bF_t) &= \vec(\widehat \bU_1^{\top} \bU_1 \bF_t \bU_2^{\top} \widehat \bU_2) +  \vec(\widehat \bU_1^{\top} \bE_t \widehat \bU_2/\lambda) \notag\\
&=\vec(\bF_t)+\left((\widehat\bU_2^{\top} \bU_2)\otimes(\widehat\bU_1^{\top} \bU_1) -I \right) \vec(\bF_t) +  \vec(\widehat \bU_1^{\top} \bE_t \widehat \bU_2/\lambda) \notag\\
&:=\vec(\bF_t) +M\vec(\bF_t)+\vec(\widetilde\bE_t)/\lambda. \label{eq:ft-hat-decomp}
\end{align}
It follows that
\begin{align}\label{eq:ft_hat}
\vec(\widehat \bF_t) - \Phi\vec(\widehat \bF_{t-1}) &=    M\vec(\bF_t)+\vec(\widetilde\bE_t)/\lambda -\Phi M\vec(\bF_{t-1})-\Phi\vec(\widetilde\bE_{t-1})/\lambda + \vec(\bxi_t).
\end{align}

For any $\bar\Phi$, we have
\begin{align*}
&\sum_{t=2}^T \| \vec(\widehat \bF_t) -\bar\Phi\vec(\widehat \bF_{t-1}) \|_2^2 - \sum_{t=2}^T  \| \vec(\widehat \bF_t) -\Phi\vec(\widehat \bF_{t-1}) \|_2^2 \\
=&\sum_{t=2}^T \left[ (\Phi-\bar\Phi) \vec(\widehat \bF_{t-1})  \right]^{\top} \left[ 2\left( \vec(\widehat\bF_t)-\Phi\vec(\widehat \bF_{t-1}) \right) +(\Phi-\bar\Phi)\vec(\widehat \bF_{t-1}) \right] \\
=&\sum_{t=2}^T \left[ (\Phi-\bar\Phi) \vec(\widehat \bF_{t-1})  \right]^{\top}  \left[ (\Phi-\bar\Phi) \vec(\widehat \bF_{t-1})  \right] + 2\sum_{t=2}^T \left[ (\Phi-\bar\Phi) \vec(\widehat \bF_{t-1})  \right]^{\top}  \left[\vec(\widehat\bF_{t}) -\Phi \vec(\widehat \bF_{t-1}) \right]\\
:=& \I+\II.
\end{align*}
We first bound I. By the ergodic theorem,
\begin{align*}
\frac1T\sum_{t=2}^T  \vec(\bF_{t-1}) \vec(\bF_{t-1})^{\top} \to \bGamma_0\qquad a.s.
\end{align*}
By \eqref{eq:ft_hat}, we can decompose I as follows
\begin{align*}
\I=&T\cdot \text{tr}\left[(\Phi-\bar\Phi)  \bGamma_0 (\Phi-\bar\Phi) ^{\top}  \right] + T\cdot \text{tr}\left[(\Phi-\bar\Phi)  \left( \frac1T \sum_{t=2}^T\vec(\bF_{t-1}) \vec(\bF_{t-1})^{\top} -\bGamma_0\right) (\Phi-\bar\Phi) ^{\top}  \right] \\
&+   2\sum_{t=2}^T \text{tr}\left[(\Phi-\bar\Phi)  (M\vec(\bF_{t-1})+\vec(\widetilde\bE_{t-1})/\lambda) \vec(\bF_{t-1})^{\top} (\Phi-\bar\Phi) ^{\top}  \right]  \\
&+ \sum_{t=2}^T \text{tr}\left[(\Phi-\bar\Phi)  (M\vec(\bF_{t-1})+\vec(\widetilde\bE_{t-1})/\lambda) (M\vec(\bF_{t-1})+\vec(\widetilde\bE_{t-1})/\lambda)^{\top} (\Phi-\bar\Phi) ^{\top}  \right]\\
:=&\I_1+ \I_2+\I_3+\I_4.
\end{align*}
By Lemma 1 in \cite{cai2018}, Corollaries 3.1 and 3.2 in \cite{han2020} imply that
\begin{align}\label{rate:tfm}
\|M\|_{\rm S}=O \left(\max_{k\le 2} \|\widehat \bU_k^{\top} \bU_k -I\|_{\rm S} \right) =O_{\P}\left( \frac{\sigma\sqrt{d_{\max}} }{\lambda\sqrt{T}} \right) =o_{\P}(1).
\end{align}
On the boundary set $\|\Phi-\bar\Phi\|_{\rm F}=\eta_T$, by calculating the variance and employing \eqref{eq:lemma-error-3} in Lemma \ref{lemma-error}, we know that
\begin{align}
\I_2&=o_{\P}(\eta_T^2 T), \label{eq1:lemma-gradient}\\
\I_3+\I_4&=O_{\P}\left(\frac{\eta_T^2 \sqrt{Td_{\max}}}{\lambda}+\frac{\eta_T^2\sigma\sqrt{Td_{\max}} }{\lambda} + \frac{\eta_T^2 \sigma^2 T}{\lambda^2} \right)  =o_{\P} \left(\eta_T^2 T \right)+O_{\P} \left( \frac{\eta_T^2 \sigma^2 T}{\lambda^2} \right).   \label{eq2:lemma-gradient}
\end{align}
In addition, on the boundary set $\|\Phi-\bar\Phi\|_{\rm F}=\eta_T$,
\begin{align}\label{eq3:lemma-gradient}
\I_1=T\cdot\text{tr}\left[(\Phi-\bar\Phi)  \bGamma_0 (\Phi-\bar\Phi) ^{\top}  \right] \ge \lambda_{\min}(\Gamma_0)\eta_T^2 T,
\end{align}
where $\lambda_{\min}(\Gamma)$ is the minimum eigenvalue of $\Gamma_0$, which is strictly positive under the condition that $\bA_1,\bA_2,\Cov(\vec(\bxi_t))$ are nonsingular.

Next, we consider to bound II. Again, by \eqref{eq:ft_hat}, we can decompose II as follows
\begin{align*}
\II=&2\sum_{t=2}^T \vec(\bF_{t-1})^{\top} (I+M^{\top})(\Phi-\bar\Phi)^{\top}  \vec(\bxi_t) +  2\sum_{t=2}^T \vec(\widetilde\bE_{t-1})^{\top} (\Phi-\bar\Phi)^{\top} \vec(\bxi_t)/\lambda \\
& + 2\sum_{t=2}^T \vec(\bF_{t-1})^{\top} (I+M^{\top})(\Phi-\bar\Phi)^{\top}  (M\vec(\bF_t)-\Phi M\vec(\bF_{t-1})) \\
& +  2\sum_{t=2}^T \vec(\widetilde\bE_{t-1})^{\top}  (\Phi-\bar\Phi)^{\top} (M\vec(\bF_t)-\Phi M\vec(\bF_{t-1}))/\lambda \\
&+ 2\sum_{t=2}^T \vec(\bF_{t-1})^{\top} (I+M^{\top})(\Phi-\bar\Phi)^{\top}  (\vec(\widetilde\bE_t)-\Phi \vec(\widetilde\bE_{t-1}))/\lambda \\
& +  2\sum_{t=2}^T \vec(\widetilde\bE_{t-1})^{\top} (\Phi-\bar\Phi)^{\top} (\vec(\widetilde\bE_t)-\Phi \vec(\widetilde\bE_{t-1}))/\lambda^2 \\
:=& \II_1+\II_2+\II_3+\II_4+\II_5+\II_6.
\end{align*}
on the boundary set $\|\Phi-\bar\Phi\|_{\rm F}=\eta_T$, elementary calculation shows that
\begin{align*}
\II_1 =O_{\P} \left(\eta_T \sqrt{T} \right),\quad \II_3 =O_{\P}\left( \frac{\eta_T\sqrt{Td_{\max}}}{\lambda}\right),\quad \II_6=O_{\P}\left(\frac{\eta_T \sigma^2 T}{\lambda^2} \right) .
\end{align*}
On the other hand, on the boundary set $\|\Phi-\bar\Phi\|_{\rm F}=\eta_T$, \eqref{eq:lemma-error-3} and \eqref{eq:lemma-error-4} in Lemma \ref{lemma-error} imply that
\begin{align*}
\II_2=O_{\P} \left( \frac{\eta_T \sigma\sqrt{Td_{\max}}}{\lambda} \right), \quad \II_4=O_{\P} \left( \frac{ \eta_T\sigma^2 d_{\max} }{ \lambda^2} \right), \quad \II_5=O_{\P} \left( \frac{ \eta_T\sigma \sqrt{Td_{\max}} }{ \lambda} \right) .
\end{align*}
It follows that
\begin{align}\label{eq4:lemma-gradient}
\II=O_{\P}\left(\eta_T\sqrt{T}+\frac{\eta_T\sigma \sqrt{Td_{\max}}}{\lambda} + \frac{\eta_T \sigma^2 T}{\lambda^2} \right).
\end{align}
Note that $\I\ge 0$. Combining \eqref{eq1:lemma-gradient}-\eqref{eq4:lemma-gradient}, and with \eqref{eq:lemma-gradient1}, we have
\begin{align}\label{eq5:lemma-gradient}
\P\left( \inf_{\|\widebar\Phi-\Phi\|_{\rm F}= \eta_T} \sum_{t=2}^T \left\| \vec(\widehat \bF_t) -\widebar\Phi \vec(\widehat \bF_{t-1}) \right\|_2^2 \le \sum_{t=2}^T \left\| \vec(\widehat \bF_t) -\Phi \vec(\widehat \bF_{t-1}) \right\|_2^2 \right)\to 0    .
\end{align}
Since $\sum_{t=2}^T \left\| \vec(\widehat \bF_t) -\widebar\Phi \vec(\widehat \bF_{t-1}) \right\|_2^2$ is a convex function of $\bar\Phi$, \eqref{eq:lemma-gradient2} is implied by \eqref{eq5:lemma-gradient} and the convexity.
\end{proof}

\begin{proof}[\bf Proof of Proposition \ref{prop:factor}]
As $\widehat \bF_t=\widehat \bU_1^{\top} \bX_t\widehat\bU_2/\lambda$, write
\begin{align*}
\widehat \bF_t&=\bR_1 \bF_t \bR_2^\top + \widehat\bU_1^{\top}(\bU_1-\widehat \bU_1 \bR_1)\bF_t\bU_2^{\top}\widehat\bU_2 + \bR_1 \bF_t(\bU_2-\widehat\bU_2 \bR_2)^\top \widehat\bU_2 + \widehat \bU_1^{\top} \bE_t\widehat\bU_2/\lambda.
\end{align*}
By Lemma 1 in \cite{cai2018}, Corollaries 3.1 (for iTOPUP) and 3.2 (for iTIPUP) in \cite{han2020} imply that for certain rotation matrices $\bR_1,\bR_2$,
\begin{align*}
\max_{k\le 2} \|\bU_k -\widehat \bU_k \bR_k \|_{\rm S}  =O_{\P}\left( \frac{\sigma\sqrt{d_{\max}} }{\lambda\sqrt{T}} \right) .
\end{align*}
It follows that
\begin{align*}
\|\widehat\bF_t -\bR_1 \bF_t \bR_2^\top \|_{\rm S} =  O_{\P}\left( \frac{\sigma\sqrt{d_{\max}} }{\lambda\sqrt{T}} \right)+    \widehat \bU_1^{\top} \bE_t\widehat\bU_2/\lambda = O_{\P}\left( \frac{\sigma\sqrt{d_{\max}} }{\lambda\sqrt{T}} + \frac{\sigma}{\lambda} \right) .
\end{align*}
Next, we prove \eqref{eq:prop:factor1}. For notational simplicity, assume $\bR_1=I_{r_1}$ and $\bR_2=I_{r_2}$. Write
\begin{align*}
\vec(\widehat \bF_t) &= \vec(\widehat \bU_1^{\top} \bU_1 \bF_t \bU_2^{\top} \widehat \bU_2) +  \vec(\widehat \bU_1^{\top} \bE_t \widehat \bU_2/\lambda) \\
&=\vec(\bF_t)+\left((\widehat\bU_2^{\top} \bU_2)\otimes(\widehat\bU_1^{\top} \bU_1) -I \right) \vec(\bF_t) +  \vec(\widehat \bU_1^{\top} \bE_t \widehat \bU_2/\lambda) \\
&:=\vec(\bF_t) +M\vec(\bF_t)+\vec(\widetilde\bE_t)/\lambda.
\end{align*}
Note that $M=O_{\P}( (\sigma\sqrt{d_{\max}} )/(\lambda\sqrt{T}) $. It follows that
\begin{align*}
&\frac{1}{T-h}\sum_{t=h+1}^T  \vec(\widehat \bF_{t}) \vec(\widehat \bF_{t-h})^\top  - \frac{1}{T-h}\sum_{t=h+1}^T  \vec( \bF_{t}) \vec(\bF_{t-h})^\top \\
=&\frac{1}{T-h}\sum_{t=h+1}^T  \vec(\bF_{t}) \vec(\bF_{t-h})^\top M^\top + \frac{1}{T-h}\sum_{t=h+1}^T  M\vec(\bF_{t}) \vec(\bF_{t-h})^\top   \\
&+ \frac{1}{T-h}\sum_{t=h+1}^T  M \vec(\bF_{t}) \vec(\bF_{t-h})^\top M^\top + \frac{1}{T-h}\sum_{t=h+1}^T  (I+M)\vec(\bF_{t}) \vec(\widetilde\bE_{t-h})^\top/\lambda \\
&+ \frac{1}{T-h}\sum_{t=h+1}^T  \vec(\widetilde\bE_{t})/\lambda\cdot \vec(\bF_{t-h})^\top (I+M)^\top +\frac{1}{T-h}\sum_{t=h+1}^T  \vec(\widetilde\bE_{t}) \vec(\widetilde\bE_{t-h})^\top/\lambda^2 \\
:=& \II_1 + \II_2 + \II_3 + \II_4 + \II_5 + \II_6.
\end{align*}
By Lemma \ref{lemma-error}, we have
\begin{align*}
\II_i =    O_{\P}\left( \frac{\sigma\sqrt{d_{\max}} }{\lambda\sqrt{T}} \right),
\end{align*}
for $i=1,...,6$. Then we obtain \eqref{eq:prop:factor1}.

The last term \eqref{eq:prop:factor2} will be similar. The only difference is that now $\II_6=O_{\P}(\sigma^2/\lambda^2)$.
\end{proof}

In the following proof of main theorems, without loss of generality, we will assume $\bR_1=I_{r_1}$ and $\bR_2=I_{r_2}$.

\begin{proof}[\bf Proof of Theorem \ref{thm:a}]
Without loss of generality, assume $\bR_1=I_{r_1}$ and $\bR_2=I_{r_2}$.
Let $\cS_m=\{M:M\in\R^{m\times m},\|M\|_{\rm F}=1\}$. Since $\bA_1,\bA_2$ are nonsingular and $\bA_1\in\cS_{r_1}$, it can be showed that if $\widetilde \bA_1\in\cS_{r_1}$ and $\| \widetilde\bA_1 - \bA_1\|_{\rm F}^2+\| \widetilde\bA_2 - \bA_2\|_{\rm F}^2\ge \eta_T^2$, then $\| (\widetilde \bA_2 \otimes\widetilde\bA_1) - (\bA_2\otimes \bA_1)\|_{\rm F}^2\ge C \eta_T^2$, where $C$ is a constant depending on $\bA_1$ and $\bA_2$. By Lemma \ref{lemma-gradient},
\begin{align*}
\P\left( \inf_{\| \widetilde\bA_1 - \bA_1\|_{\rm F}^2+\| \widetilde\bA_2 - \bA_2\|_{\rm F}^2\ge \eta_T^2} \sum_{t=2}^T \left\| \vec(\widehat \bF_t) -\widebar\Phi \vec(\widehat \bF_{t-1}) \right\|_2^2 \le \sum_{t=2}^T \left\| \vec(\widehat \bF_t) -\Phi \vec(\widehat \bF_{t-1}) \right\|_2^2 \right)\to 0,
\end{align*}
with $\widetilde\bA_1\in\cS_{r_1}$, $\Phi=\bA_2\otimes\bA_1$. It follows that
\begin{align*}
\P\left(\| \widehat\bA_1 - \bA_1\|_{\rm F}^2+\| \widehat\bA_2 - \bA_2\|_{\rm F}^2\ge \eta_T^2 \right)    \to 0,
\end{align*}
with $\widehat\bA_1\in\cS_{r_1}$.
Condition \eqref{eq:lemma-gradient1} implies that
\begin{align}\label{rate:a-hat}
\|\widehat \bA_1-\bA_1\|_{\rm S}=O_{\P}(\epsilon_T),\quad \text{and}\quad \|\widehat\bA_2-\bA_2\|_{\rm S}=O_{\P}(\epsilon_T),
\end{align}
where $\epsilon_T=\sigma\sqrt{d_{\max}}/(\lambda\sqrt{T})+\sigma^2/\lambda^2+1/\sqrt{T}=O(1)$.

The gradient condition for the least square estimator of second step MAR is:
\begin{align}
\sum_{t}\widehat \bA_1 \widehat \bF_{t-1} \widehat \bA_2^{\top} \widehat \bA_2 \widehat \bF_{t-1}^{\top} -\sum_t \widehat \bF_t \widehat \bA_2 \widehat \bF_{t-1}^{\top} &=0 ,  \label{eq1:thm:a} \\
\sum_{t}\widehat \bA_2 \widehat \bF_{t-1}^{\top} \widehat \bA_1^{\top} \widehat \bA_1 \widehat \bF_{t-1} -\sum_t \widehat \bF_t^{\top} \widehat \bA_1 \widehat \bF_{t-1} &=0 . \label{eq2:thm:a}
\end{align}
As $\widehat \bF_t=\widehat \bU_1^{\top} \bX_t\widehat\bU_2/\lambda$, write
\begin{align*}
\widehat \bF_t&=\bF_t+ (\widehat\bU_1^{\top}\bU_1-I)\bF_t\bU_2^{\top}\widehat\bU_2 + \bF_t(\bU_2^{\top}\widehat\bU_2-I) + \widehat \bU_1^{\top} \bE_t\widehat\bU_2/\lambda: = \bF_t+ \widecheck \bF_t.
\end{align*}
Then, \eqref{eq1:thm:a} can be rewritten as follows
\begin{align*}
&\sum_{t}\widehat \bA_1 \bF_{t-1} \widehat \bA_2^{\top} \widehat \bA_2  \bF_{t-1}^{\top} -\sum_t \bF_t \widehat \bA_2  \bF_{t-1}^{\top} \\
=& \sum_t \widecheck \bF_t \widehat\bA_2 \bF_{t-1}^{\top}+\sum_t \bF_{t}\widehat \bA_2 \widecheck\bF_{t-1}^{\top} + \sum_{t} \widecheck\bF_t \widehat\bA_2 \widecheck\bF_{t-1}^{\top}  - \sum_{t}\widehat \bA_1 \widecheck\bF_{t-1} \widehat \bA_2^{\top} \widehat \bA_2  \bF_{t-1}^{\top} \\
&- \sum_{t}\widehat \bA_1 \bF_{t-1} \widehat \bA_2^{\top} \widehat \bA_2  \widecheck\bF_{t-1}^{\top} -\sum_{t}\widehat \bA_1 \widecheck\bF_{t-1} \widehat \bA_2^{\top} \widehat \bA_2  \widecheck\bF_{t-1}^{\top}.
\end{align*}
By \eqref{rate:tfm}, \eqref{rate:a-hat} and Lemma \ref{lemma-error}, the right hand side of the above identity satisfies
\begin{align*}
{\rm RHS}&=O_{\P}\left(\frac{\sigma\sqrt{Td_{\max}} }{ \lambda} \right) +  O_{\P}\left(\frac{\sigma^2 d_{\max}}{\lambda^2}+\frac{\sigma^2 T}{\lambda^2} \right) + O_{\P}\left(\frac{\sigma\sqrt{Td_{\max}} }{ \lambda} \right) +  O_{\P}\left(\frac{\sigma^2 d_{\max}}{\lambda^2}+\frac{\sigma^2 T}{\lambda^2} \right)\\
&= O_{\P}\left(\frac{\sigma\sqrt{Td_{\max}} }{ \lambda} + \frac{\sigma^2 T}{\lambda^2} \right).
\end{align*}
The left hand side satisfies
\begin{align*}
{\rm LHS}&= \sum_{t}\widehat \bA_1 \bF_{t-1} \widehat \bA_2^{\top} \widehat \bA_2  \bF_{t-1}^{\top} -\sum_t \bA_1\bF_{t-1}\bA_2^{\top} \widehat \bA_2  \bF_{t-1}^{\top}  -\sum_t \bxi_t \widehat \bA_2  \bF_{t-1}^{\top} \\
&= \sum_{t}(\widehat \bA_1 -\bA_1) \bF_{t-1}  \widehat\bA_2^{\top} \widehat\bA_2  \bF_{t-1}^{\top} -\sum_t \bA_1\bF_{t-1} (\widehat\bA_2-\bA_2)^{\top} \widehat\bA_2  \bF_{t-1}^{\top}  -\sum_t \bxi_t \widehat\bA_2  \bF_{t-1}^{\top} \\
&= \sum_{t}(\widehat \bA_1 -\bA_1) \bF_{t-1}  \bA_2^{\top} \bA_2  \bF_{t-1}^{\top} -\sum_t \bA_1\bF_{t-1} (\widehat\bA_2-\bA_2)^{\top} \bA_2  \bF_{t-1}^{\top}  -\sum_t \bxi_t \bA_2  \bF_{t-1}^{\top} +O_{\P}(\epsilon_T^2 T+\epsilon_T\sqrt{T}).
\end{align*}
Thus, from \eqref{eq1:thm:a}, as $\epsilon_T\gtrsim 1/\sqrt{T}$, we have
\begin{align}\label{eq3:thm:a}
&\sum_{t}(\widehat \bA_1 -\bA_1) \bF_{t-1}  \bA_2^{\top} \bA_2  \bF_{t-1}^{\top} -\sum_t \bA_1\bF_{t-1} (\widehat\bA_2-\bA_2)^{\top} \bA_2  \bF_{t-1}^{\top} \notag\\
&=\sum_t \bxi_t \bA_2  \bF_{t-1}^{\top} +  O_{\P}\left(\frac{\sigma\sqrt{Td_{\max}} }{ \lambda} + \frac{\sigma^2 T}{\lambda^2} + \epsilon_T^2 T\right)    .
\end{align}
Similarly, from \eqref{eq2:thm:a}, by \eqref{rate:tfm}, \eqref{rate:a-hat} and Lemma \ref{lemma-error}, we can derive
\begin{align}\label{eq4:thm:a}
&\sum_{t}(\widehat \bA_2 -\bA_2) \bF_{t-1}^{\top}  \bA_1^{\top} \bA_1  \bF_{t-1} -\sum_t \bA_2\bF_{t-1}^{\top} (\widehat\bA_1-\bA_1)^{\top} \bA_1  \bF_{t-1} \notag \\
&=\sum_t \bxi_t^{\top} \bA_1  \bF_{t-1} + O_{\P}\left(\frac{\sigma\sqrt{Td_{\max}} }{ \lambda} + \frac{\sigma^2 T}{\lambda^2} + \epsilon_T^2 T\right)    .
\end{align}
Hence, combing \eqref{eq3:thm:a} and \eqref{eq4:thm:a}, we have
\begin{align*}
&\begin{pmatrix} (\sum_{t}\bF_{t-1}  \bA_2^{\top} \bA_2  \bF_{t-1}^{\top})\otimes \bI_{r_1} & \sum_t (\bF_{t-1}\bA_2^{\top})\otimes (\bA_1\bF_{t-1})\\ \sum_t (\bA_2 \bF_{t-1}^{\top})\otimes (\bF_{t-1}^{\top} \bA_1^{\top}) & \bI_{r_2} \otimes (\sum_t \bF_{t-1}^{\top} \bA_1^{\top} \bA_1 \bF_{t-1})\end{pmatrix}
\begin{pmatrix}
\vec(\widehat \bA_1 -\bA_1)\\ \vec(\widehat \bA_2^{\top} -\bA_2^{\top} )
\end{pmatrix} \\
&=\sum_t \begin{pmatrix}
(\bF_{t-1}\bA_2^{\top}) \otimes \bI_{r_1} \\ \bI_{r_2} \otimes (\bF_{t-1}^{\top} \bA_1^{\top})
\end{pmatrix}
\vec(\bxi_t)+ O_{\P}\left(\frac{\sigma\sqrt{Td_{\max}} }{ \lambda} + \frac{\sigma^2 T}{\lambda^2} + \epsilon_T^2 T\right)    .
\end{align*}
Define $\bW_t^{\top}=[(\bA_2\bF_t^{\top}) \otimes \bI_{r_1}; \bI_{r_2} \otimes (\bA_1\bF_t)]\in\R^{r_1r_2\times (r_1^2+r_2^2)}$, $\bgamma=(\vec(\bA_1)^{\top},{\bf0}^{\top})^{\top} \in\R^{r_1^2+r_2^2}$ and $\bH_1:=\E(\bW_t \bW_t^{\top})+\bgamma\bgamma^{\top}$. It follows that,
\begin{align}\label{eq:thm:a0}
(\sum_t \bW_{t-1}\bW_{t-1}^{\top})
\begin{pmatrix}
\vec(\widehat \bA_1 -\bA_1)\\ \vec(\widehat \bA_2^{\top} -\bA_2^{\top} )
\end{pmatrix}
=\sum_t \bW_{t-1}\vec(\bxi_t) + O_{\P}\left(\frac{\sigma\sqrt{Td_{\max}} }{ \lambda} + \frac{\sigma^2 T}{\lambda^2} + \epsilon_T^2 T\right)    .
\end{align}
By the ergodic theorem,
\begin{align*}
\frac1T\sum_{t=2}^T  \bW_{t-1} \bW_{t-1}^{\top} \to \E\ \bW_{t-1} \bW_{t-1}^{\top} \qquad a.s.
\end{align*}

Let $\balpha_2:=\vec(\bA_2^\top)$. Observe that $\bW_{t-1}^{\top}(\balpha_1^\top,-\balpha_2^\top)^\top=\boldsymbol{0}$, hence $\E(\bW_{t-1} \bW_{t-1}^{\top})$ is not a full rank matrix. On the other hand, since we require $\|\bA_1\|_{\rm F}=1$ and $\|\widehat\bA_1\|_{\rm F}=1$, it holds that $\balpha^\top(\vec(\widehat\bA_1)-\balpha)=o_{\P}(T^{-1/2})+O_{\P}(\sigma\sqrt{d_{\max}}/(\lambda\sqrt{T})+\sigma^2/\lambda^2)$. Therefore,
\begin{align}\label{eq:thm:a}
\begin{pmatrix}
\vec(\widehat \bA_1 -\bA_1)\\ \vec(\widehat \bA_2^{\top} -\bA_2^{\top} )
\end{pmatrix}
=\bH_1^{-1}\frac1T \sum_{t=2}^T \bW_{t-1}\vec(\bxi_t) +  O_{\P}\left(\frac{\sigma\sqrt{d_{\max}} }{ \lambda\sqrt{T}} + \frac{\sigma^2 }{\lambda^2} + \epsilon_T^2 \right) =O_{\P}(\epsilon_T ) .
\end{align}
\end{proof}

\begin{proof}[\bf Proof of Theorem \ref{thm:b}]
As $\lambda/\sigma\gg \sqrt{d_{\max}}+T^{1/4}$, \eqref{eq:thm:a} implies that
\begin{align*}
\begin{pmatrix}
\vec(\widehat \bA_1 -\bA_1)\\ \vec(\widehat \bA_2^{\top} -\bA_2^{\top} )
\end{pmatrix}
=\bH_1^{-1}\frac{1}{T} \sum_{t=2}^T \bW_{t-1}\vec(\bxi_t) + o_{\P}\left(\frac{1 }{ \sqrt{T}} \right).
\end{align*}
By martingale central limit theorem,
\begin{align*}
\frac{1}{\sqrt{T}} \sum_{t=2}^T \bW_{t-1}\vec(\bxi_t)   \to N(0,\E(\bW_{t-1}\Sigma \bW_{t-1}^{\top})).
\end{align*}
Then the desired central limit theorem follows.
\end{proof}

Now, let us consider the lagged least square estimator.
The ideal optimization problem in \eqref{eq:optim2} is equivalent to
\begin{align*}
\min\limits_{\bA_1,\bA_2} \sum\limits_t \| \vec(\bF_t) - (\bA_2 \otimes \bA_1) \vec(\widetilde\bF_{t-2}) \|_{\rm F}^2,
\end{align*}
where $\widetilde\bGamma_k=T^{-1}\sum_t\vec(\bF_t)\vec(\bF_{t-k})^{\top}$ and $\vec(\widetilde\bF_t)=\widetilde\bGamma_1\widetilde\bGamma_0^{-1}\vec(\bF_t)$. Then, we can reformulate the above optimization problem as
\begin{align*}
\min\limits_{\bA_1,\bA_2} \sum\limits_t \| \bF_t - \bA_1\widetilde\bF_{t-2}\bA_2^{\top} \|_{\rm F}^2.
\end{align*}
Let $\vec(\bG_t)=\bGamma_1\bGamma_0^{-1}\vec(\bF_t)\in\R^{r_1r_2}$ with $\bGamma_k=\E\vec(\bF_t)\vec(\bF_{t-k})^{\top}\in\R^{r_1r_2\times r_1r_2}$. 
Then $\vec(\bG_t)=(\bA_2\otimes \bA_1)\vec(\bF_t)$ and $\bG_t=\bA_1 \bF_t \bA_2^\top$. The ideal optimization problem of lagged estimator becomes
\begin{align}\label{eq:ideal_lagopt}
\min\limits_{\bA_1,\bA_2} \sum\limits_t \| \bF_t - \bA_1\bG_{t-2}\bA_2^{\top} \|_{\rm F}^2.
\end{align}

\begin{lemma}\label{lemma-gradient2}
Let $\widehat\bGamma_k=T^{-1}\sum_{t=3}^T \vec(\widehat \bF_t)\vec(\widehat \bF_{t-k})$, $\bGamma_k=\E \vec(\bF_t)\vec(\bF_{t-k})$ for $k=0,1,2$, and $\vec(\widehat\bG_t)=\widehat\bGamma_1\widehat\bGamma_0^{-1}\vec(\widehat\bF_t)$.
Suppose Assumptions \ref{asmp:error} and \ref{asmp:mar-noise} holds. Let $\Phi=\bA_2\otimes \bA_1$, $d_{\max}=\max\{d_1,d_2\}$. For any sequence $\{\eta_T\}$ such that
\begin{align}\label{eq:lemma-gradient3}
\frac{\eta_T}{\sigma\sqrt{d_{\max}}/(\lambda\sqrt{T})+(\sigma^2/\lambda^2+1)/\sqrt{T}} \to \infty,
\end{align}
then
\begin{align}\label{eq:lemma-gradient4}
\P\left(\inf_{\|\widebar\Phi-\Phi\|_{\rm F}\ge \eta_T} \sum_{t=3}^T \left\| \vec(\widehat \bF_t) -\widebar\Phi \vec(\widehat \bG_{t-2}) \right\|_2^2 \le \sum_{t=3}^T \left\| \vec(\widehat \bF_t) -\Phi \vec(\widehat \bG_{t-2}) \right\|_2^2 \right)\to 0    .
\end{align}
\end{lemma}

\begin{proof}
For any $\bar\Phi$, we have
\begin{align*}
&\sum_{t=3}^T \| \vec(\widehat \bF_t) -\bar\Phi\vec(\widehat \bG_{t-2}) \|_2^2 - \sum_{t=2}^T  \| \vec(\widehat \bF_t) -\Phi\vec(\widehat \bG_{t-2}) \|_2^2 \\
=&\sum_{t=3}^T \left[ (\Phi-\bar\Phi) \vec(\widehat \bG_{t-2})  \right]^{\top} \left[ 2\left( \vec(\widehat\bF_t)-\Phi\vec(\widehat \bG_{t-2}) \right) +(\Phi-\bar\Phi)\vec(\widehat \bG_{t-2}) \right] \\
=&\sum_{t=3}^T \left[ (\Phi-\bar\Phi) \vec(\widehat \bG_{t-2})  \right]^{\top}  \left[ (\Phi-\bar\Phi) \vec(\widehat \bG_{t-2})  \right] + 2\sum_{t=3}^T \left[ (\Phi-\bar\Phi) \vec(\widehat \bG_{t-2})  \right]^{\top}  \left[\vec(\widehat\bF_{t}) -\Phi \vec(\widehat \bG_{t-2}) \right]\\
:=& \I+\II.
\end{align*}
Applying the arguments in Lemma \ref{lemma-gradient}, we can bound I in a similar way and obtain the same rates.

Next, we bound II. As $\bGamma_2=\Phi\bGamma_1$,
\begin{align*}
\II=&2\sum_{t=3}^T \vec(\widehat\bG_{t-2})^{\top} (\Phi-\bar\Phi)^{\top}  [\vec(\widehat\bF_t) -\Phi \vec(\widehat\bG_{t-2}) ]\\
=&2\sum_{t=3}^T \vec(\widehat\bF_{t-2})^{\top} \widehat\bGamma_0^{-1}\widehat\bGamma_1^{\top}(\Phi-\bar\Phi)^{\top}  [\vec(\widehat\bF_t) -\Phi \widehat\bGamma_1\widehat\bGamma_0^{-1} \vec(\widehat\bF_{t-2}) ] \\
=&2\tr\Big\{\sum_{t=3}^T \widehat\bGamma_0^{-1}\widehat\bGamma_1^{\top}(\Phi-\bar\Phi)^{\top}  [\vec(\widehat\bF_t) -\Phi \widehat\bGamma_1\widehat\bGamma_0^{-1} \vec(\widehat\bF_{t-2}) ]\vec(\widehat\bF_{t-2})^{\top} \Big\} \\
=&2T\tr \Big\{\widehat\bGamma_0^{-1}\widehat\bGamma_1^{\top}(\Phi-\bar\Phi)^{\top}  [\widehat\bGamma_2 -\Phi \widehat\bGamma_1\widehat\bGamma_0^{-1} \widehat\bGamma_{0} ] \Big\} \\
=&2T\tr \Big\{\widehat\bGamma_0^{-1}\widehat\bGamma_1^{\top}(\Phi-\bar\Phi)^{\top}  [\widehat\bGamma_2 -\bGamma_2 -\Phi (\widehat\bGamma_1-\bGamma_1) ] \Big\}
\end{align*}
Elementary calculation shows that
\begin{align}\label{eq:gammak}
\widehat\bGamma_k-\bGamma_k =O_{\P} \left(\frac{\sigma\sqrt{d_{\max}}}{\lambda\sqrt{T}} + \frac{\sigma^2}{\lambda^2\sqrt{T}} + \frac{1}{\sqrt{T}} \right),\quad k=1,2.
\end{align}
It follows that
\begin{align}\label{eq4:lemma-gradient2}
\II=O_{\P}\left(\eta_T\sqrt{T}+\frac{\eta_T\sigma \sqrt{Td_{\max}}}{\lambda} + \frac{\eta_T \sigma^2 \sqrt{T}}{\lambda^2} \right).
\end{align}
Note that $\I\ge 0$. Then we have
\begin{align}\label{eq5:lemma-gradient2}
\P\left( \inf_{\|\widebar\Phi-\Phi\|_{\rm F}= \eta_T} \sum_{t=3}^T \left\| \vec(\widehat \bF_t) -\widebar\Phi \vec(\widehat \bG_{t-2}) \right\|_2^2 \le \sum_{t=3}^T \left\| \vec(\widehat \bF_t) -\Phi \vec(\widehat \bG_{t-2}) \right\|_2^2 \right)\to 0    .
\end{align}
Since $\sum_{t=3}^T \left\| \vec(\widehat \bF_t) -\widebar\Phi \vec(\widehat \bG_{t-2}) \right\|_2^2$ is a convex function of $\bar\Phi$, \eqref{eq:lemma-gradient4} is implied by \eqref{eq5:lemma-gradient2} and the convexity.
\end{proof}

\begin{proof}[\bf Proof of Theorem \ref{thm:laga}]
Without loss of generality, assume $\bR_1=I_{r_1}$ and $\bR_2=I_{r_2}$.
Let $\Phi=\bA_2\otimes\bA_1$, $\widetilde\Phi=\widehat\bGamma_2\widehat\bGamma_1^{-1}$, $\widehat\Omega=\widehat\bGamma_1\widehat\bGamma_0^{-1}\widehat\bGamma_1^\top$ and $\epsilon_T=\sigma\sqrt{d_{\max}}/(\lambda\sqrt{T})+\sigma^2/(\lambda^2\sqrt{T})=o(1)$. By the ergodic theorem,
\begin{align*}
\widetilde\bGamma_1=\frac1T\sum_{t=2}^T\vec(\bF_t)\vec(\bF_{t-1})\to\Phi\bGamma_0, \quad a.s.
\end{align*}
By \eqref{eq:gammak} and using similar arguments to those in the proof of Part II in Lemma \ref{lemma-gradient},
\begin{align}\label{eq:phit}
\widetilde\Phi &=\Phi+\frac1T\sum_{t=3}^T \vec(\bxi_t)\vec(\bF_{t-2})\widehat\bGamma_1^{-1}+O_{\P}\left(\frac{\sigma\sqrt{d_{\max}}}{\lambda\sqrt{T}} + \frac{\sigma^2}{\lambda^2\sqrt{T}} \right)   \notag\\
&= \Phi+\frac1T\sum_{t=3}^T \vec(\bxi_t)\vec(\bF_{t-2})\bGamma_1^{-1}+O_{\P}\left(\epsilon_T \right) +o_{\P}(T^{-1/2}) \notag\\
&:= \Phi +\theta_T +O_{\P}\left(\epsilon_T \right) +o_{\P}(T^{-1/2}),
\end{align}
where $\theta_T=O_{\P}(T^{-1/2})$ and $\theta_T\in\R^{r_1r_2\times r_1r_2}$. By martingale central limit theorem, we have
\begin{align*}
\sqrt{T}\vec(\theta_T) \to N(0,(\bGamma_1^{-\top}\bGamma_0\bGamma_1^{-1} )\otimes \Sigma)  .
\end{align*}
Let $\widehat\Omega^{1/2} = (\widehat\bGamma_1\widehat\bGamma_0^{-1}\widehat\bGamma_1^\top )^{-1/2}=(\widehat\bw_1,...,\widehat\bw_{r_1r_2})$ and $\widehat\omega_i=\mat_1(\widehat\bw_i)\in\R^{r_1\times r_2}$, $1\le i\le r_1r_2$.
Then $\Omega^{1/2}=(\bGamma_1 \bGamma_0^{-1} \bGamma_1^\top)^{1/2}=(\Phi \bGamma_0 \Phi)^{1/2}$.
Similarly, we can show
\begin{align}\label{eq:wi}
\widehat \bw_i=\bw_i+O_{\P}(\epsilon_T+T^{-1/2}).
\end{align}

The optimization problem for lagged estimator \eqref{eq:optim2} is equivalent to
\begin{align}\label{eq:optim2b}
(\widehat\bA_1,\widehat\bA_2) &= \argmin\limits_{\bA_1,\bA_2} \tr\left( (\bA_2 \otimes \bA_1 - \widetilde\Phi)\widehat\Omega (\bA_2 \otimes \bA_1 - \widetilde\Phi)^\top \right) \notag \\
&= \argmin\limits_{\bA_1,\bA_2} \sum_{i=1}^{r_1 r_2}\| (\bA_2 \otimes \bA_1 - \widetilde\Phi) \widehat\bw_i \|_2^2.
\end{align}
Then, the gradient condition for the lagged least square estimator is:
\begin{align}
\sum_{i=1}^{r_1r_2}\widehat\bA_1 \widehat\omega_i \widehat \bA_2^\top \widehat\bA_2 \widehat\omega_i^\top - \sum_{i=1}^{r_1r_2}\mat_1(\widetilde\Phi \widehat\bw_i) \widehat\bA_2 \widehat\omega_i^\top &=0 ,  \label{eq1:thm:laga} \\
\sum_{i=1}^{r_1r_2}\widehat\bA_2 \widehat\omega_i^\top \widehat \bA_1^\top \widehat\bA_1 \widehat\omega_i - \sum_{i=1}^{r_1r_2}\mat_1^\top (\widetilde\Phi \widehat\bw_i) \widehat\bA_1 \widehat\omega_i &=0 . \label{eq2:thm:laga}
\end{align}
Similar to the proof of Theorem \ref{thm:a}, condition \eqref{eq:lemma-gradient3} in Lemma \ref{lemma-gradient2} implies that
\begin{align}\label{rate:laga-hat}
\|\widehat \bA_1-\bA_1\|_{\rm S}=O_{\P}(\epsilon_T+T^{-1/2}),\quad \text{and}\quad \|\widehat\bA_2-\bA_2\|_{\rm S}=O_{\P}(\epsilon_T+T^{-1/2}).
\end{align}
Employing \eqref{eq:phit}, \eqref{eq:wi}, \eqref{rate:laga-hat}, the gradient condition can be rewritten as follows
\begin{align*}
& \sum_{i=1}^{r_1r_2}(\widehat\bA_1 -\bA_1)\omega_i \bA_2^\top \bA_2 \omega_i^\top +\sum_{i=1}^{r_1 r_2} \bA_1 \omega_i (\widehat\bA_2 -\bA_2)^\top \bA_2 \omega_i^\top = \sum_{i=1}^{r_1r_2}\mat_1(\theta_T \bw_i) \bA_2 \omega_i^\top + O_{\P}\left(\epsilon_T \right) +o_{\P}(T^{-1/2}) , \\
& \sum_{i=1}^{r_1r_2}\omega_i^\top \bA_1^\top (\widehat\bA_1 - \bA_1) \omega_i \bA_2^\top +\sum_{i=1}^{r_1 r_2} \omega_i^\top \bA_1^\top \bA_1 \omega_i (\widehat\bA_2 -\bA_2)^\top = \sum_{i=1}^{r_1r_2}\omega_i^\top \bA_1^\top\mat_1(\theta_T \bw_i) + O_{\P}\left(\epsilon_T \right) +o_{\P}(T^{-1/2}) .
\end{align*}
Hence, we have
\begin{align}\label{eq:clt-lag-1}
&\begin{pmatrix} (\sum_{i=1}^{r_1r_2}\omega_i  \bA_2^{\top} \bA_2  \omega_i^{\top})\otimes \bI_{r_1} & \sum_{i=1}^{r_1r_2} (\omega_i\bA_2^{\top})\otimes (\bA_1\omega_i)\\ \sum_{i=1}^{r_1r_2} (\bA_2 \omega_i^{\top})\otimes (\omega_i^{\top} \bA_1^{\top}) & \bI_{r_2} \otimes (\sum_{i=1}^{r_1r_2} \omega_i^{\top} \bA_1^{\top} \bA_1 \omega_i)\end{pmatrix}
\begin{pmatrix}
\vec(\widehat \bA_1 -\bA_1)\\ \vec(\widehat \bA_2^{\top} -\bA_2^{\top} )
\end{pmatrix} \notag\\
&=\sum_{i=1}^{r_1r_2} \begin{pmatrix}
(\omega_i\bA_2^{\top}) \otimes \bI_{r_1} \\ \bI_{r_2} \otimes (\omega_i^{\top} \bA_1^{\top})
\end{pmatrix}
\theta_T \bw_i+ O_{\P}\left(\epsilon_T \right) +o_{\P}(T^{-1/2})    .
\end{align}

Recall the LSE estimator in Theorems \ref{thm:a} and \ref{thm:b} can be reformulated as
\begin{align}\label{eq:optim3b}
(\widehat\bA_1^{(\rm LSE)},\widehat\bA_2^{(\rm LSE)}) &= \argmin\limits_{\bA_1,\bA_2} \sum_{i=1}^{r_1 r_2} \| (\bA_2 \otimes \bA_1 - \widehat\bGamma_1) \widehat\bpsi_i \|_{\rm F}^2,
\end{align}
where $\widehat\bGamma_0^{-1/2}=(\widehat\bpsi_1,...,\widehat\bpsi_{r_1r_2})$ and $\phi_i=\mat_1(\bpsi_i)$.
The above arguments implies that
\begin{align}\label{eq:clt-lag-2}
&\begin{pmatrix} (\sum_{i=1}^{r_1r_2}\phi_i  \bA_2^{\top} \bA_2  \phi_i^{\top})\otimes \bI_{r_1} & \sum_{i=1}^{r_1r_2} (\phi_i\bA_2^{\top})\otimes (\bA_1\phi_i)\\ \sum_{i=1}^{r_1r_2} (\bA_2 \phi_i^{\top})\otimes (\phi_i^{\top} \bA_1^{\top}) & \bI_{r_2} \otimes (\sum_{i=1}^{r_1r_2} \phi_i^{\top} \bA_1^{\top} \bA_1 \phi_i)\end{pmatrix}
\begin{pmatrix}
\vec(\widehat \bA_1^{(\rm LSE)} -\bA_1)\\ \vec(\widehat \bA_2^{(\rm LSE)\top} -\bA_2^{\top} )
\end{pmatrix} \notag\\
&=\sum_{i=1}^{r_1r_2} \begin{pmatrix}
(\phi_i\bA_2^{\top}) \otimes \bI_{r_1} \\ \bI_{r_2} \otimes (\phi_i^{\top} \bA_1^{\top})
\end{pmatrix}
\frac1T\sum_{t=2}^T \vec(\bxi_t)\vec(\bF_{t-1})\bGamma_0^{-1} \bpsi_i+ O_{\P}\left(\frac{\sigma\sqrt{d_{\max}} }{\lambda\sqrt{T} }+\frac{\sigma^2}{\lambda^2} \right) +o_{\P}(T^{-1/2})    .
\end{align}
From comparing \eqref{eq:clt-lag-1} with \eqref{eq:clt-lag-2}, and \eqref{eq:optim2b} with \eqref{eq:optim3b}, we arrive at a result akin to \eqref{eq:thm:a0}:
\begin{align}\label{eq:thm:a1}
(\sum_t \bQ_{t-2}\bQ_{t-2}^{\top})
\begin{pmatrix}
\vec(\widehat \bA_1 -\bA_1)\\ \vec(\widehat \bA_2^{\top} -\bA_2^{\top} )
\end{pmatrix}
=\sum_t \bQ_{t-2}\vec(\bxi_t) + O_{\P}\left(\epsilon_T \right) +o_{\P}(T^{-1/2})    .
\end{align}
where $\bQ_t^{\top}=[(\bA_2\bG_t^{\top}) \otimes \bI_{r_1}; \bI_{r_2} \otimes (\bA_1\bG_t)]\in\R^{r_1r_2\times (r_1^2+r_2^2)}$, $\bG_t=\bA_1 \bF_t \bA_2^\top$, $\bgamma=(\vec(\bA_1)^{\top},{\bf0}^{\top})^{\top} \in\R^{r_1^2+r_2^2}$ and $\bH_2:=\E(\bQ_t \bQ_t^{\top})+\bgamma\bgamma^{\top}$.

Consequently, analogous to \eqref{eq:thm:a}, we have that
\begin{align}\label{eq:thm:lag}
\begin{pmatrix}
\vec(\widehat \bA_1 -\bA_1)\\ \vec(\widehat \bA_2^{\top} -\bA_2^{\top} )
\end{pmatrix}
=\bH_2^{-1}\frac1T \sum_{t=2}^T \bQ_{t-2}\vec(\bxi_t) +  O_{\P}\left(\epsilon_T \right) +o_{\P}(T^{-1/2}) = O_{\P}\left(\frac{\sigma\sqrt{d_{\max}} }{ \lambda\sqrt{T}} + \frac{\sigma^2 }{\lambda^2\sqrt{T}} + \frac{1}{\sqrt{T}} \right) .
\end{align}

\end{proof}

\begin{proof}[\bf Proof of Theorem \ref{thm:lagb}]
As $\lambda/\sigma\gg \sqrt{d_{\max}}$, \eqref{eq:thm:lag} implies that
\begin{align*}
\begin{pmatrix}
\vec(\widehat \bA_1 -\bA_1)\\ \vec(\widehat \bA_2^{\top} -\bA_2^{\top} )
\end{pmatrix}
=\bH_2^{-1}\frac1T \sum_{t=2}^T \bQ_{t-2}\vec(\bxi_t) + o_{\P}\left(\frac{1 }{ \sqrt{T}} \right).
\end{align*}
By martingale central limit theorem,
\begin{align*}
\frac{1}{\sqrt{T}} \sum_{t=2}^T \bQ_{t-2}\vec(\bxi_t)   \to N\left(0,\E \bQ_t\Sigma \bQ_t^\top \right).
\end{align*}
Then the desired central limit theorem follows.
\end{proof}

\begin{proof}[\bf Proof of Proposition~\ref{prop:efficiency}.] First of all, observe that the asymptotic distributions established in both Theorem~\ref{thm:b} and Theorem~\ref{thm:lagb} are as if there were no estimation errors in $\hat\bF_t$. Therefore, it suffices to consider the estimation of \eqref{eq:dmfm2} based on observed $\bF_t$. The lagged estimator is obtained through the minimization
\begin{align*}
    \min\limits_{\bA_1,\bA_2} \| (\bA_2 \otimes \bA_1 - \hat{\bGamma}_2 \hat{\bGamma}_1^{-1}) (\hat{\bGamma}_1 \hat{\bGamma}_0^{-1} \hat{\bGamma}_1^\top)^{1/2} \|_{\rm F}^2.
\end{align*}
An inspection of the proofs of Theorems \ref{thm:b} and \ref{thm:lagb} reveal that the minimizer of
\begin{equation}\label{eq:lag_0}
    \min\limits_{\bA_1,\bA_2} \| (\bA_2 \otimes \bA_1 - \hat{\bGamma}_2 {\bGamma}_1^{-1}) ({\bGamma}_1 {\bGamma}_0^{-1}{\bGamma}_1^\top)^{1/2} \|_{\rm F}^2
\end{equation}
have the same asymptotic distribution.

Now we introduce a new model as the instrument of the proof. Consider the matrix regression model
\begin{equation}\label{eq:mat_reg}
    \bY_t = \bA_1\bZ_t\bA_2^\top+\bxi_t,
\end{equation}
where $\bZ_t\stackrel{d}{=}\bA_1\bF_t\bA_2^\top$, $\bxi_t$ are same as those in \eqref{eq:dmfm2}, and $\{\bZ_t\}$ and $\{\bxi_t\}$ are independent. Let $\bY'=[\vec(\bY_1),\ldots,\vec(\bY_T)]$ and $\bZ'=[\vec(\bZ_1),\ldots,\vec(\bZ_T)]$, and set $\bGamma_{yz}=\E[\vec(\bY_t)\vec(\bZ_t)^\top]$, $\bGamma_{zz}=\E[\vec(\bZ_t)\vec(\bZ_t)^\top]$, $\hat\bGamma_{yz}=T^{-1}\bY'\bZ$ and $\hat\bGamma_{zz}=T^{-1}\bZ'\bZ$. The least squares estimation of \eqref{eq:mat_reg} is given by
\begin{equation}\label{eq:reg_lse}
    \min\limits_{\bA_1,\bA_2} \sum_{t=1}^T\|\bY_t - \bA_1\bZ_t\bA_2^\top\|_{\mathrm F}^2,
\end{equation}
which is equivalent to
\begin{equation*}
    \min\limits_{\bA_1,\bA_2} \| (\bA_2 \otimes \bA_1 - \hat{\bGamma}_{yz} \hat{\bGamma}_{zz}^{-1}) \hat\bGamma_{zz}^{1/2} \|_{\rm F}^2.
\end{equation*}
The asymptotic distribution of the minimizers above is the same as those from
\begin{equation}\label{eq:reg_0}
    \min\limits_{\bA_1,\bA_2} \| (\bA_2 \otimes \bA_1 - \hat{\bGamma}_{yz} {\bGamma}_{zz}^{-1}) \bGamma_{zz}^{1/2} \|_{\rm F}^2,
\end{equation}
{with $\hat{\bGamma}_{zz}$ replaced by $\bGamma_{zz}$.} Note that the asymptotic distribution of $\hat\bGamma_2\bGamma_1^{-1}$ in \eqref{eq:lag_0} is
\begin{equation*}
    \sqrt{T}\vec\left( \hat\bGamma_2\bGamma_1^{-1} - \bA_2\otimes\bA_1 \right) \Rightarrow N\left[\boldsymbol{0},(\bGamma_1^{-\top}\bGamma_0\bGamma_1^{-1})\otimes\Sigma\right],
\end{equation*}
and the asymptotic distribution of $\hat\bGamma_{yz}\bGamma_{zz}^{-1}$ in \eqref{eq:reg_0} is
\begin{equation*}
    \sqrt{T}\vec\left( \hat\bGamma_{yz}\bGamma_{zz}^{-1} - \bA_2\otimes\bA_1 \right) \Rightarrow N\left(\boldsymbol{0},\bGamma_{zz}^{-1}\otimes\Sigma\right).
\end{equation*}
Since $\bZ_t\stackrel{d}{=}\bA_1\bF_t\bA_2^\top$, it follows that $\bGamma_{zz}=(\bA_2\otimes\bA_1)\bGamma_0(\bA_2^\top\otimes\bA_1^\top)=\bGamma_1\bGamma_0^{-1}\bGamma_1^\top$. Therefore, the preceding observations imply that the minimizers in \eqref{eq:lag_0} and \eqref{eq:reg_0} have the same asymptotic distribution, and hence the lagged estimator and the LSE in \eqref{eq:reg_lse} have the same asymptotic distribution.

When $\Sigma=\sigma_{\xi}^2\bI$, by the proof of Theorem~\ref{thm:b},
\begin{equation*}
    \Xi_1= \sigma_{\xi}^2\left(\E(\bW_t\bW_t^\top)+\bgamma\bgamma^\top\right)^{-1}\E(\bW_t\bW_t^\top)\left(\E(\bW_t\bW_t^\top)+\bgamma\bgamma^\top\right)^{-1},
\end{equation*}
where $\bW_t^{\top}=[(\bA_2\bF_t^{\top}) \otimes \bI; \bI \otimes (\bA_1\bF_t)]$.
From \eqref{eq:reg_lse}, a similar argument shows that
\begin{equation*}
    \Xi_2= \sigma_{\xi}^2\left(\E(\tilde\bW_t\tilde\bW_t^\top)+\bgamma\bgamma^\top\right)^{-1}\E(\tilde\bW_t\tilde\bW_t^\top)\left(\E(\tilde\bW_t\tilde\bW_t^\top)+\bgamma\bgamma^\top\right)^{-1},
\end{equation*}
where $\tilde\bW_t^{\top}=[(\bA_2\bZ_t^{\top}) \otimes \bI; \bI \otimes (\bA_1\bZ_t)]$.
Let $\check\bW_t^{\top}=[(\bA_2\bxi_t^{\top}) \otimes \bI; \bI \otimes (\bA_1\bxi_t)]$. Since $\bZ_t\stackrel{d}{=}\bA_1\bF_t\bA_2^\top$ and $\bF_t=\bA_1\bF_{t-1}\bA_2^\top+\bxi_t$, it follows that
\begin{equation}\label{eq:eff}
  \E(\bW_t\bW_t^\top) = \E(\tilde\bW_t\tilde\bW_t^\top) + \E(\check\bW_t\check\bW_t^\top)\succeq \E(\tilde\bW_t\tilde\bW_t^\top).
\end{equation}

\newcommand{\bzero}{\boldsymbol{0}}
To continue to show that $\Xi_1\preceq\Xi_2$,
we first note that the expression of $\Xi_i$ would be the same if $\bgamma$ is replaced by any multiple $c\bgamma$ with $c\neq 0$. Let $\balpha_1=\vec(\bA_1)$ and $\balpha_2=\vec(\bA_2^\top)$. Note that $\|\balpha_1\|=1$. In the following we replace $\bgamma$ by
\begin{align*}
    \bar\bgamma & :={\sqrt{1+\|\balpha_2\|^2}}\,\bgamma={\sqrt{1+\|\balpha_2\|^2}}\begin{pmatrix}
        \balpha_1 \\ \bzero
    \end{pmatrix} = \frac{1}{\sqrt{1+\|\balpha_2\|^2}}
    \begin{pmatrix}
    \balpha_1\\ -\balpha_2
    \end{pmatrix}
    + \frac{1}{\sqrt{1+\|\balpha_2\|^2}}
    \begin{pmatrix}
    \|\balpha_2\|^2\balpha_1 \\ \balpha_2
    \end{pmatrix} \\
    & =:\bgamma_0+\bgamma_1,
\end{align*}
such that $\bgamma_0$ is a unit vector and $\bgamma_1^\top\bgamma_0=\bzero$.
Let $\bH=\E(\bW_t\bW_t^\top)$. Note that $\bH$ is rank difficient by one, and $\bH\bgamma_0=\bzero$. Let $\bH^{+}$ be the Moore-Penrose inverse of $\bH$. Multiple applications of the Woodbury matrix identity shows that
\begin{equation*}
    \left(\bH + \bar\bgamma\bar\bgamma^\top\right)^{-1} = \bH^{+} + (1+\bgamma_1^\top\bH^{+}\bgamma_1)\bgamma_0\bgamma_0^\top -\bgamma_0\bgamma_1^\top\bH^{+} - \bH^{+}\bgamma_1\bgamma_0^\top.
\end{equation*}
It follows that $\left(\bH + \bar\bgamma\bar\bgamma^\top\right)^{-1}\bar\bgamma = \bgamma_0$, and therefore
\begin{equation*}
    \Xi_1 = \left(\bH + \bar\bgamma\bar\bgamma^\top\right)^{-1} - \left(\bH + \bar\bgamma\bar\bgamma^\top\right)^{-1} \bar\bgamma\bar\bgamma^\top\left(\bH + \bar\bgamma\bar\bgamma^\top\right)^{-1} = \left(\E(\bW_t\bW_t^\top)+\bar\bgamma\bar\bgamma^\top\right)^{-1} - \bgamma_0\bgamma_0^\top.
\end{equation*}
Similarly,
\begin{equation*}
    \Xi_2  = \left(\E(\tilde\bW_t\tilde\bW_t^\top)+\bar\bgamma\bar\bgamma^\top\right)^{-1} - \bgamma_0\bgamma_0^\top.
\end{equation*}
Now it is evident that $\Xi_1\preceq\Xi_2$ in view of \eqref{eq:eff}.
\end{proof}

\end{document}